\documentclass[a4paper,11pt]{article}
\usepackage[T1]{fontenc}
\usepackage{framed}
\usepackage{a4wide}

\usepackage{stmaryrd}
\usepackage[pdftex]{graphicx}
\usepackage[table]{xcolor}
\usepackage{caption}
\usepackage{longtable}
\usepackage{amsmath,amssymb, subcaption}
\usepackage{hyperref}
\hypersetup{hidelinks}
\usepackage{parskip}

\usepackage{tikz}
\usetikzlibrary{fit,positioning,arrows.meta,tikzmark}

\usepackage{enumerate}

\providecommand{\cebwt}{\textsc{ceBWT}}
\providecommand{\ctmatch}{\approx_{\mathrm{CT}}}
\providecommand{\cteq}{=_{\omega}}
\providecommand{\ctprec}{\prec_{\omega}}
\providecommand{\ctpeq}{\preceq_{\omega}}
\providecommand{\ctree}[1]{\mathrm{ct}(#1)}
\providecommand{\pde}[1]{\langle #1\rangle}
\providecommand{\rpde}[1]{\langle #1\rangle_{\mathrm r}}
\providecommand{\rts}[1]{\mathopen{[\![}#1\mathclose{]\!]}_{\mathrm r}}
\providecommand{\CA}[1]{\mathrm{CA}_{#1}}
\providecommand{\ICA}[1]{\mathrm{CA}_{#1}^{-1}}
\providecommand{\LF}[1]{\mathrm{LF}_{#1}}
\providecommand{\FL}[1]{\mathrm{FL}_{#1}}
\providecommand{\Larr}[1]{\mathrm{L}_{#1}}
\providecommand{\Farr}[1]{\mathrm{F}_{#1}}
\providecommand{\LCPinf}[1]{\mathrm{LCP}^{\infty}_{#1}}
\providecommand{\rankop}{\operatorname{rank}}
\providecommand{\selectop}{\operatorname{select}}
\providecommand{\rot}[2]{\operatorname{rot}(#1,#2)}
\providecommand{\conj}[2]{\operatorname{conj}_{#1}(#2)}
\providecommand{\crange}[2]{\operatorname{CR}_{#1}(#2)}
\providecommand{\lcpct}[2]{\operatorname{lcp}^{\infty}(#1,#2)}
\providecommand{\primitive}[1]{\operatorname{root}(#1)}
\providecommand{\rta}[2]{\operatorname{Rot}(#1,#2)}
\providecommand{\cnt}[2]{\operatorname{cnt}_{#1}(#2)}
\providecommand{\plcp}[2]{\operatorname{plcp}^{\infty}_{#1}(#2)}
\providecommand{\slcp}[2]{\operatorname{slcp}^{\infty}_{#1}(#2)}
\providecommand{\cntarr}[1]{\operatorname{cnt}[#1]}
\providecommand{\plcparr}[1]{\operatorname{plcp}^{\infty}[#1]}
\providecommand{\slcparr}[1]{\operatorname{slcp}^{\infty}[#1]}

\definecolor{mygreen}{RGB}{0,120,0}
\definecolor{myorange}{RGB}{200,120,0}
\definecolor{mygray}{gray}{0.7}
\providecommand{\one}{\cellcolor{myorange}\textcolor{white}{1}}
\providecommand{\zero}{\cellcolor{mygreen}\textcolor{white}{0}}
\providecommand{\emptyc}{\cellcolor{mygray}}
\usepackage{nicefrac}
\newcommand{\twoFold}{\textsf{\ensuremath{\nicefrac{2}{3}}-fold}}

\usepackage[capitalise]{cleveref}

\newtheorem{lemma}{Lemma}
\providecommand{\wurz}[1]{\operatorname{root}(#1)}

\title{Compact multi-text index for circular Cartesian tree matching\thanks{This work is based on the Bachelor thesis of the first author~\cite{baPauliRoman}.}}
\date{}
\author{Roman Pauli \and Eric Osterkamp \and Dominik K\"{o}ppl}
\begin{document}
\maketitle
\begin{abstract}
Cartesian tree matching (CTM) is a structural pattern matching approach that identifies sequences with the same Cartesian tree topology, making it suitable for data with natural variability where exact comparisons carry little semantic meaning.
While theoretical algorithms for CTM have been studied extensively, systematic empirical evaluations of practical implementations remain rare.
This article presents an implementation of the Cartesian Extended Burrows--Wheeler Transform (ceBWT), a BWT-based index structure for CTM.
The implementation supports both a dynamically extendable and a statically compressed index variant.
Two extensions to the original construction are provided: an alternative to fourfold replication that uses a combination of threefold (for adding new sequences) and twofold (for the initial text) input replication for improved computational efficiency, called \twoFold{} and multithreaded support.
The implementation is evaluated on genomic and MIDI datasets across varying sequence lengths, sequence counts, and pattern lengths.
The results show that both index variants scale consistently with the expected asymptotic trends.
The static index achieves competitive index sizes and fast pattern search, and it has lower query time than the included order-preserving index on the evaluated MIDI workload while solving a weaker structural matching problem and requiring more memory and time during construction.
The dynamic index introduces higher construction overhead due to the complexity of incremental merging, which restricts practical scalability to smaller sequences.
Repetitiveness has a measurable but modest effect on memory consumption and construction speed, but a strong effect on match frequency.
Compared to KMP-based CTM baselines, the ceBWT index offers significantly faster repeated pattern search when an index is already available.
Overall, this work provides a first systematic empirical analysis of a ceBWT-based CTM implementation, identifying both its practical strengths and limitations, and establishing a foundation for future optimization.
\end{abstract}


\textbf{Source Code: } \url{https://github.com/koeppl/cebwt}

\section{Introduction}
\label{chapter:introduction}

String matching is a long-standing field of computer science. 
One important structural variant of exact string matching is \emph{order-preserving matching} (OPM), 
which compares relative order rather than absolute values~\cite{kim14orderpreserving, kim_representations_2017}. 
Efficient methods for OPM are already available~\cite{chhabra_filtration_2016, cho15fast, kim_order-preserving_2023}.

This article focuses on \emph{Cartesian tree matching} (CTM), which compares the Cartesian tree topology of two sequences~\cite{vuillemin80unifying,park19cartesiantreematching}. In contrast to OPM, CTM abstracts away pairwise rank comparisons and instead preserves the recursive structure induced by local and global extrema~\cite{park20patternsperiods}. This behavior is useful when natural variability or measurement noise makes exact comparisons less meaningful, for example in music, audio, and genomic data, where structural shape can be more informative than exact values~\cite{kim14orderpreserving,meysman_dna_2012}.

\subsection{Problem and Contributions}
Although CTM has been studied in several theoretical variants~\cite{park19cartesiantreematching, kim21compact, kim_approximate_2025}, practical implementations with systematic measurements of runtime, memory use, and scalability remain limited. The Cartesian Extended Burrows--Wheeler Transform (ceBWT)~\cite{osterkamp26cebwt} provides a suitable foundation for such an evaluation because it supports both a dynamically extendable index and a compact static index for circular CTM.

This article asks how a ceBWT-based CTM implementation performs in construction time, search time, memory consumption, and scalability, and how it compares with exact string-matching indexes, an order-preserving FM-index, and KMP-based CTM baselines. To answer this question, we implement the ceBWT, add multithreaded construction and reduced input replication, and evaluate the resulting dynamic and static variants on genomic and MIDI datasets across different sequence lengths, sequence counts, and pattern lengths. The contribution is a practical implementation and systematic benchmark that identifies the strengths, limitations, and implementation challenges of ceBWT-based CTM.

\subsection{Structure}
\cref{sec:technical-background} introduces the theoretical foundations needed to understand the implementation. \cref{chapter:StateOfArt} summarizes the state of the art. \cref{section:softwarearchitecture} describes implementation details and design decisions. \cref{chapter:methodic} presents the datasets and benchmark methodology. \cref{chapter:benchmarks} presents and discusses the benchmark results. \cref{sec:conclusion-outlook} answers the research questions and outlines future work. 

\subsection{State of the Art}
\label{chapter:StateOfArt}
CTM was first introduced by Park et al.~\cite{park19cartesiantreematching} in 2019. Their work defines both the matching model and an algorithm for pattern search. The approach can be implemented using parent-distance encoding (PDE) or the Cartesian tree signature. Single-pattern search follows a KMP-like~\cite{knuth77kmp} approach: a failure function is computed during pattern preprocessing and encodes the longest proper prefix of the pattern's Cartesian tree that is also a suffix. During the search phase, this failure function avoids redundant comparisons by skipping already matched prefixes after a mismatch. CTM can be applied to single-pattern and multiple-pattern matching. For a single pattern of length $m$ in a text of length $n$, the search can be completed in $\mathcal{O}(m+n)$ time. For multiple patterns, let $k$ denote the number of patterns and $m$ their total length. The search can be solved deterministically in $\mathcal{O}((m+n)\,\log k)$ time or in randomized $\mathcal{O}(m+n)$ time. The construction of the index requires randomized $\mathcal{O}(n)$ time.
In this approach, indexing and pattern search are directly connected and cannot be separated.

An extension of this approach was proposed in~\cite{kim21compact}, focusing on the construction of a compact index that is independent of the pattern. Instead of PDE, this index uses the Cartesian tree signature~\cite{demaine14cartesian}. Although this encoding was already discussed by Park et al.~\cite{park19cartesiantreematching}, the compact-index construction implements it in a more space-efficient way. The index is built on the $L$ and $F$ arrays described in~\cite[Section 4.1]{kim21compact},
which store values related to the Cartesian tree signature in Burrows--Wheeler order and enable LF mapping over Cartesian tree suffixes.
Using these two sets of bit vectors, values can be stored efficiently without additional structures. Pattern search is performed by backward search over these structures, analogously to the FM-index~\cite{ferragina00fmindex}, without requiring the pattern during index construction. This approach constructs a permanent index structure using $3n+o(n)$ bits and supports pattern search in $\mathcal{O}(m)$ time. Unlike previous approaches, the index structure is independent of the patterns and can therefore be used for arbitrary patterns.

This work builds on the ceBWT~\cite{osterkamp26cebwt}, which further extends CTM. It supports dynamic extension by merging newly indexed sequences into an existing index and updating the $L$, $F$, and $\mathrm{LCP}^{\infty}$ structures accordingly. It can also be converted to a compact static representation, which loses dynamic update support but matches the space usage of the previous compact approach. In addition, ceBWT indexes circular rotations rather than only the original sequences. This is achieved by means of a conjugate array, which considers all rotations of each sequence during construction and maps each rotation to its canonical representative. Pattern search follows a backward-search approach analogous to~\cite{kim21compact} for both the dynamic and static indexes.
The index can be constructed in $\mathcal{O}(n \frac{\lg\,\sigma\,\lg\,n}{\lg\,\lg\,n})$ time using $\mathcal{O}(n\,\lg\,\sigma)$ bits of memory, where $\sigma$ is the alphabet size. Its static form requires $3n+o(n)$ bits. Pattern search can be performed in $\mathcal{O}(m \frac{\lg\,\sigma\,\lg\,n}{\lg\,\lg\,n})$ time for the dynamic index and $\mathcal{O}(m)$ time for the static index.

We also point out other related works that are not directly based on the ceBWT but are relevant to the CTM problem.
First, Kim and Han~\cite{kim_approximate_2025} present an alternative method for verifying CTM, introducing a verification method that tolerates bounded mismatches.
Second, Song et al.~\cite{song21fastmatchingalg} provide fast algorithms for single and multiple pattern CTM, including SIMD-based solutions for short patterns and filtration-based approaches for multiple patterns.

In contrast to the works above, which focus primarily on theoretical complexity and algorithmic design, this article provides an empirical evaluation of the ceBWT. Runtime, memory usage, and scalability are systematically assessed across different input sizes and sequence counts, and the results are compared against established exact string matching algorithms, the compact order-preserving FM-index of Decaroli et al.~\cite{decaroli19compact}, and two KMP-based CTM variants based on~\cite{park19cartesiantreematching}.

\section{Technical Background}\label{sec:technical-background}\label{chapter:kap2}
This section fixes the notation and the exact objects implemented by our index.
We use one-based indexing.
For integers $i\le j$, $[i..j]$ denotes the set $\{i,i+1,\ldots,j\}$ and is empty if $j<i$.
For a string $X$, $X[i]$ is the $i$-th symbol, $X[i..j]$ is the corresponding substring, $|X|$ is its length, and $X^\omega$ is the infinite periodic string obtained by repeating $X$.
All strings are over a finite totally ordered alphabet $\Sigma$.
When convenient, we identify $\Sigma$ with an integer interval~$[1..\sigma]$; the particular
choice of interval is immaterial. For CTM and OPM, order-isomorphic
renumberings preserve all matches, and for exact matching a uniform renumbering
preserves equality. Hence, an input alphabet such as the MIDI pitch range
$[0..127]$ can equivalently be shifted to $[1..128]$ for notational purposes.
For constructions that need a separator, we use a symbol $\$\notin\Sigma$ smaller than every symbol of $\Sigma$.
After such a renumbering, lexicographic comparisons of PDEs use the order $\$ < 1 < \cdots < \sigma < \infty$,
where $\infty$ is a special symbol larger than every symbol of $\Sigma$ and $\$$.
\cref{tab:notation-quick-reference} in the appendix gives a quick lookup for the main notation used throughout the paper.
For examples, we write the contents of a string over an integer alphabet in the form $X = \\llbracket x_1, x_2, \ldots, x_n \\rrbracket$, where $x_i = X[i]$.

\subsection{Order-preserving matching (OPM)}
OPM compares the relative order of values rather than their actual magnitudes~\cite{kim14orderpreserving,kim_representations_2017}.
It is mentioned here only as a nearby matching model: two strings match in the order-preserving sense if all pairwise order relations among their positions agree.
CTM is coarser in the sense relevant for this paper, because it keeps the recursive minimum structure of the sequence instead of the full rank order.

\subsection{Cartesian tree matching (CTM)}\label{section:CTM}\label{sec:ctm-definitions}
The Cartesian tree $\ctree{X}$ of a string $X\in(\Sigma\cup\{\$\})^*$ is defined recursively.
The empty string has the empty Cartesian tree.
For a nonempty string $X$, let $i$ be the leftmost position of the minimum symbol in $X$.
Then $i$ is the root, $\ctree{X[1..i-1]}$ is the left subtree, and $\ctree{X[i+1..|X|]}$ is the right subtree.
Two strings $X$ and $Y$ Cartesian-tree match, written $X\ctmatch Y$, if their Cartesian trees are identical.
Thus the values themselves are not matched; only the topology induced by the positions of recursively chosen minima is matched.

\begin{figure}[t]
    \centering
    \begin{subfigure}[b]{0.45\textwidth}
        \centering
        \fbox{\begin{tikzpicture}
\node at (1.5, 0) {\text{i)}};
            \node (r) at (2.5, 0) {3};
            \node (l1) at (2, -0.8) {5};
            \node (r1) at (3, -0.8) {4};
            \node (l2) at (2.5, -1.6) {9};
            \node (r2) at (3.5, -1.6) {5};
            \draw (r) -- (l1);
            \draw (r) -- (r1);
            \draw (l1) -- (l2);
            \draw (r1) -- (r2);
        
\node at (0.5, -2.4) {\text{ii)}};
            \node (r3) at (1, -2.4) {3};
            \node (l3) at (0.5, -3.2) {9};
            \node (r4) at (1.5, -3.2) {4};
            \draw (r3) -- (l3);
            \draw (r3) -- (r4);
        
\node at (4, -2.4) {\text{iii)}};
            \node (r5) at (4.5, -2.4) {1};
            \node (l5) at (4, -3.2) {2};
            \node (r6) at (5, -3.2) {3};
            \draw (r5) -- (l5);
            \draw (r5) -- (r6);
        \end{tikzpicture}}
        \caption{CTM}
        \label{fig:CTM}
    \end{subfigure}
    \hfill
    \begin{subfigure}[b]{0.45\textwidth}
        \centering
        \fbox{\begin{tabular}{l r@{: }l}
                \text{i)}   & $S$          & $\llbracket 5, 9, 3, 4, 5\rrbracket$ \\
                            & $R(S)$       & $\llbracket 3, 5, 1, 2, 4\rrbracket$ \\[6pt]
                \text{ii)}  & $S[2..4]$    & $\llbracket 9, 3, 4\rrbracket$ \\
                            & $R(S[2..4])$ & $\llbracket 3, 1, 2\rrbracket$ \\[6pt]
                \text{iii)} & $P$          & $\llbracket 2, 1, 3\rrbracket$ \\
                            & $R(P)$       & $\llbracket 2, 1, 3\rrbracket$ \\
            \end{tabular}}
        \caption{OPM}
        \label{fig:OPM}
    \end{subfigure}
    \caption{Pattern matching for CTM and OPM using the sequence $S=\llbracket 5, 9, 3, 4, 5\rrbracket $, the pattern $P=\llbracket 2, 1, 3\rrbracket $, and the interval $[2..4] $.
    }
    \label{fig:OPMVSCTM}
\end{figure}
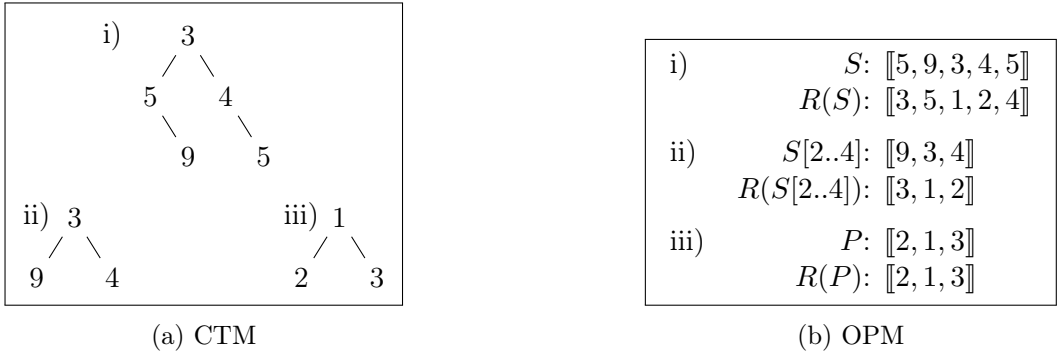

In contrast to OPM, CTM does not retain the full pairwise order of values. Instead, it focuses on the structural shape of the sequence, which is determined by the positions of recursively selected extrema.
This approach is particularly useful for pattern matching scenarios in which the local peaks of a sequence are more important than the exact ordering of values. \cref{fig:OPMVSCTM} demonstrates this difference using a concrete example.
First, the Cartesian tree of the entire sequence $S$ is shown in (i), followed by the Cartesian tree of the subsequence $S_2=S[2..4]$ in (ii), which matches the pattern in (iii). The Cartesian trees of the subsequence and the pattern have the same structure. The same subsequence does not match under OPM. In CTM, the fact that $S_2[2]=3$ is smaller than its neighbors is more important than comparing those neighbors with each other. In contrast, for OPM, the fact that $S_2[1]=9$ is larger than $S_2[3]=4$ leads to a different ranking from the pattern 
where $P[1]=2 < 3 = P[3]$.

\subsection{Cartesian tree encodings}\label{section:CTS}
A convenient representation is the \emph{parent-distance encoding} (\emph{PDE}).
Let $\infty$ be a symbol larger than every integer.
For $X\in(\Sigma\cup\{\$\})^*$, define $\pde{X}$ by
\[
\pde{X}[i]=
\begin{cases}
 i-\max\{j<i\mid X[j]\le X[i]\}, & \text{if }X[i]\ne\$\text{ and such a }j\text{ exists},\\
 X[i], & \text{if }X[i]=\$,\\
 \infty, & \text{otherwise.}
\end{cases}
\]
For strings of equal length, $X\ctmatch Y$ if and only if $\pde{X}=\pde{Y}$~\cite{park20patternsperiods}.
This equivalence is the basis for validation in our implementation and for defining the order used by the index.
For example,
\[
\pde{41327\$3}=
\llbracket
\infty,\infty,1,2,1,\$,1
\rrbracket
.
\]
See also \cref{fig:pde}.

The Cartesian tree signature (CTS) is the alternative encoding used for index navigation~\cite{demaine14cartesian,kim21compact}.
To avoid confusion with the ceBWT array $L$, we denote the CTS of a single sequence $S$ by $\lambda_S$.
For a sequence $S[1..n]$, the signature is an array $\lambda_S[1..n]$ in which $\lambda_S[i]$ is the number of elements popped from the stack when $S[i]$ is processed by the standard left-to-right stack construction of the Cartesian tree.
Equivalently, the signature can be written as the bit string $1^{\lambda_S[1]}01^{\lambda_S[2]}0\cdots 1^{\lambda_S[n]}0$, whose length is less than $2n$.
For example, for $S=(2,7,5,6,4,3,1)$ we obtain $\lambda_S=(0,0,1,0,2,1,2)$ and the bit string $0010011010110$.
The ceBWT implementation uses the equivalent rotational form described below, because it yields the $L$ and $F$ arrays needed for LF mapping.

The same signature can also be used for KMP-like CTM.
For this purpose, one stores an auxiliary array $D[1..n]$ together with $\lambda_S$.
If $S[i]$ is later popped from the stack by $S[j]$, then $D[i]=j-i$; otherwise $D[i]=0$.
When deleting the first symbol of the current window, one removes $\lambda_S[1]$ and $D[1]$; if $D[1]>0$, the value $\lambda_S[D[1]+1]$ is decremented before the removal.
This maintains the CTS of the shifted window, and the KMP failure function can be computed analogously.
The experiments therefore distinguish a PDE-based KMP baseline from a CTS-based KMP baseline.

\begin{figure}[t]
    \centering
    \fbox{\begin{tikzpicture}
            \tikzset{
                mybox/.style={
                    rectangle,
                    align=left,
                    inner sep=4pt
                }
            }
            
            \node[mybox, anchor=west] (s3) at (.5, -.8)  {\makebox[1.5cm][l]{$S_3$}: 
                $\llbracket 5, 9, 3, 7, 6\rrbracket$ };
            \node[mybox, anchor=west] (cts) at (.5, -1.5) {\makebox[1.5cm][l]{$\text{CTS}(S_3)$}: 
                $\llbracket 0, 0, 2, 0, 1\rrbracket$ };
            \node[draw, rectangle, fit=(s3)(cts), inner sep=4pt] {};
\node at (6, -.8) {\text{i)}};
            \node (ib_r)  at (7, -1.3)   {5};
\node at (8, -.8) {\text{ii)}};
            \node (iib_r) at (9, -1.3)   {5};
            \node (iib_r1) at (9.8, -1.8)  {9};
            \draw (iib_r) -- (iib_r1);
\node at (1, -2.5) {\text{iii)}};
            \node (iiib_r) at (2, -3.0)   {3};
            \node (iiib_l1) at (1.3, -3.8)  {5};
            \node (iiib_r2) at (2, -4.6)  {9};
            \draw (iiib_r) -- (iiib_l1);
            \draw (iiib_l1) -- (iiib_r2);
\node at (3.5, -2.5) {\text{iv)}};
            \node (iv_r) at (4.5, -3,0)   {3};
            \node (iva_l1) at (3.8, -3.8)  {5};
            \node (iva_r2) at (4.5, -4.6)  {9};
            \node (ivb_r1) at (5.2, -3.8)  {7};
            \draw (iv_r) -- (iva_l1);
            \draw (iva_l1) -- (iva_r2);
            \draw (iv_r) -- (ivb_r1);
\node at (6, -2.5) {\text{v)}};
            \node (v_r) at (7, -3,0)   {3};
            \node (va_l1) at (6.2, -3.8)  {5};
            \node (va_r2) at (6.7, -4.6)  {9};
            \node (vb_r1) at (7.8, -3.8)  {6};
            \node (vb_l2) at (7.3, -4.6)  {7};
            \draw (v_r) -- (va_l1);
            \draw (va_l1) -- (va_r2);
            \draw (v_r) -- (vb_r1);
            \draw (vb_r1) -- (vb_l2);
        \end{tikzpicture}
    }
    \caption{Cartesian tree signature $\mathrm{CTS}(S_3)$ for sequence $S_3 = [5, 9, 3, 7, 6]$.}
    \label{fig:cts}
\end{figure}
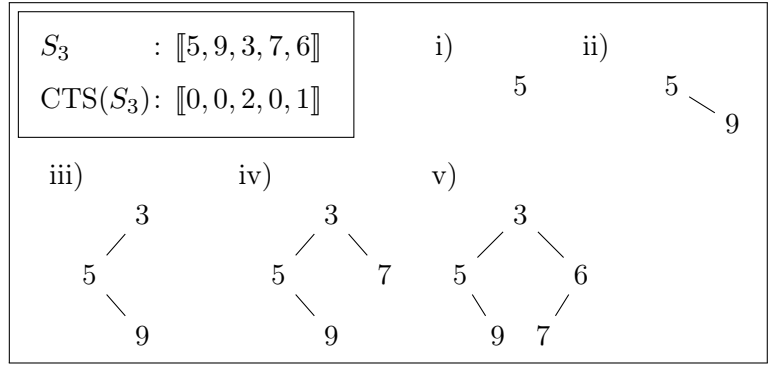

This subsection defines the Cartesian tree encodings used later in the paper.
\cref{fig:cts} shows that inserting 3 in (iii) and 6 in (v) requires restructuring to maintain the correct Cartesian tree. 
In (iii), two nodes are removed, whereas in (v), only one node is removed. For all other insertions, the value is 0 because no nodes need to be removed. Since $S_4$ from \cref{fig:pde} has the same structure, the construction process is the same. Two Cartesian tree signatures are identical if and only if the corresponding sequences yield the same removal counts.

\begin{figure}[t]
    \centering
    \fbox{\begin{tikzpicture}
            \tikzset{
                mybox/.style={
                    rectangle,
                    align=left,
                    inner sep=4pt
                }
            }
\node at (0, -0.5) {\text{ia)}};
            \node[mybox, anchor=west] (s3) at (.5, -.8)  {\makebox[1.5cm][l]{$S_3$}: [5, 9, 3, 7, 6]};
            \node[mybox, anchor=west] (pde3) at (.5, -1.6) {\makebox[1.5cm][l]{$\text{PDE}(S_3)$}: [$\infty$, 1, $\infty$, 1, 2]};
            \draw[->] (s3.south) ++(0.45, .15) to[bend left=90] ++(-0.4, 0);
            \draw[->] (s3.south) ++(1.3, .15) to[bend left=90] ++(-0.4, 0);
            \draw[->] (s3.south) ++(1.75, .15) to[bend left=110] ++(-0.85, 0);
            \node[draw, rectangle, fit=(s3)(pde3), inner sep=4pt] {};
\node at (6, -0.5) {\text{ib)}};
            \node (ib_r)  at (7.5, -0.5)   {3};
            \node (ib_l1) at (6.5, -1.3)  {5};
            \node (ib_r1) at (8.5, -1.3)  {6};
            \node (ib_l2) at (7.2, -2.1)  {9};
            \node (ib_r2) at (7.8, -2.1)  {7};
            \draw (ib_r) -- (ib_l1);
            \draw (ib_r) -- (ib_r1);
            \draw (ib_l1) -- (ib_l2);
            \draw (ib_r1) -- (ib_r2);
\node at (0, -2.6) {\text{iia)}};
            \node[mybox, anchor=west] (s4) at (.5, -3.1) {\makebox[1.5cm][l]{$S_4$}: [3, 4, 1, 8, 5]};
            \node[mybox, anchor=west] (pde4) at (.5, -3.8) {\makebox[1.5cm][l]{$\text{PDE}(S_4)$}: [$\infty$, 1, $\infty$, 1, 2]};
            \draw[->] (s3.south) ++(0.45, -2.2) to[bend left=90] ++(-0.4, 0);
            \draw[->] (s3.south) ++(1.3, -2.2) to[bend left=90] ++(-0.4, 0);
            \draw[->] (s3.south) ++(1.75, -2.2) to[bend left=110] ++(-0.85, 0);
            \node[draw, rectangle, fit=(s4)(pde4), inner sep=4pt] {};
\node at (6, -2.6) {\text{iib)}};
            \node (iib_r)  at (7.5, -2.8)  {1};
            \node (iib_l1) at (6.5, -3.6)  {3};
            \node (iib_r1) at (8.5, -3.6)  {5};
            \node (iib_l2) at (7.2,  -4.2)  {4};
            \node (iib_r2) at (7.8,  -4.2)  {8};
            \draw (iib_r) -- (iib_l1);
            \draw (iib_r) -- (iib_r1);
            \draw (iib_l1) -- (iib_l2);
            \draw (iib_r1) -- (iib_r2);
        \end{tikzpicture}
    }
    \caption{PDE $\pde{S_3}$ and $\pde{S_4}$ for sequences $S_3 = \llbracket 5, 9, 3, 7, 6\rrbracket $ and $S_4 = \llbracket 3, 4, 1, 8, 5\rrbracket $, together with the corresponding Cartesian trees.}
    \label{fig:pde}
\end{figure}
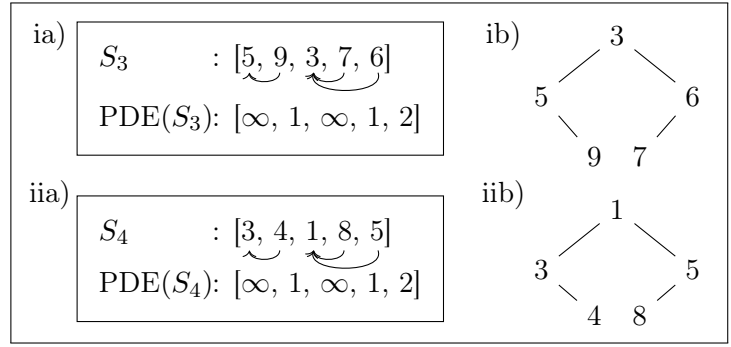

\subsection{Circular Search Space and the Count Query}\label{sec:ceBWT-query}\label{section:CEBWT}
For $X\in\Sigma^+$, let $\rot{X}{0}=X$ and $\rot{X}{k+1}=\rot{X}{k}[2..|X|]\,\rot{X}{k}[1]$.
The strings $\rot{X}{0},\ldots,\rot{X}{|X|-1}$ are the conjugates of $X$.
Given a nonempty indexed collection of texts $\mathcal T=(T_1,\ldots,T_d)$ with $T_j\in\Sigma^+$, let
\[
 n=|T_1\cdots T_d|,\qquad n_k=|T_k|.
\]
The collection notation is intentional: equal strings may occur more than once and are then indexed as distinct texts.
We identify each conjugate with its starting position in the concatenation $T_1\cdots T_d$.
For $q\in[1..n]$, let $j$ be the smallest index with $\sum_{k=1}^j n_k\ge q$ and define
\[
\conj{\mathcal T}{q}=\rot{T_j}{q-1-\sum_{k=1}^{j-1}n_k}.
\]
For a pattern $P\in\Sigma^+$, the fundamental query supported by the index is
\[
\operatorname{count}(P)=
\left|\left\{q\in[1..n]\mid P\ctmatch \conj{\mathcal T}{q}^{\omega}[1..|P|]\right\}\right|.
\]
Thus a pattern is compared against every circular rotation of every indexed text, and a match may wrap around the end of a text.
The empty pattern, if allowed by an interface, has count $n$.
The implementation additionally supports storage-dependent operations such as locating, extracting, decoding and removing indexed strings; these operations are defined precisely in \cref{section:softwarearchitecture}.

\subsection{The Order of Conjugates}\label{sec:omega-order}
The \cebwt{} index stores the conjugates in an order that makes all matches of a pattern form one interval.
For $X\in(\Sigma\cup\{\$\})^+$, define the rotational PDE
\[
\rpde{X}=\pde{X^2}[|X|+1..2|X|].
\]
Let $\primitive{Z}$ denote the primitive root of a nonempty string $Z$, i.e. the shortest string whose repetition gives $Z$.
For $X,Y\in(\Sigma\cup\{\$\})^+$, define the $\omega$-preorder $X\ctpeq Y$ if either the PDEs of the infinite strings differ and, for some prefix length $i$ reaching their first difference,
\[
\pde{X^\omega[1..i]} < \pde{Y^\omega[1..i]},
\]
or the two strings are periodic-equivalent in the sense that
\[
\primitive{\rpde{X}}=\primitive{\rpde{Y}}.
\]
We write $X\cteq Y$ if both $X\ctpeq Y$ and $Y\ctpeq X$ hold, and $X\ctprec Y$ if $X\ctpeq Y$ but not $X\cteq Y$.
This relation is a total preorder, not necessarily a total order; the implementation breaks remaining ties deterministically so that the conjugate array is a permutation.

The conjugate array $\CA{\mathcal T}[1..n]$ lists all conjugates in nondecreasing $\omega$-preorder.
The conjugate range of a pattern $P$ is the maximal interval
\[
\crange{\mathcal T}{P}=[\ell..r]
\]
such that $P\ctmatch \conj{\mathcal T}{\CA{\mathcal T}[i]}^\omega[1..|P|]$ for all $i\in[\ell..r]$.
The count query returns $0$ if this interval is empty and otherwise $r-\ell+1$.
For $P=\varepsilon$, the range is $[1..n]$.
The central purpose of the \cebwt{} index is to compute this interval by backward search without scanning the texts.

\subsection{Rotational Tree Signatures and LF Mapping}\label{sec:lf-definition}
The LF mapping moves from a conjugate to the conjugate that starts one position earlier in the same text cycle.
For each text position $q$ belonging to text $T_j$, define
\[
\operatorname{prev}_{\mathcal T}(q)=
\begin{cases}
q-1+|\primitive{\rpde{T_j}}|, & \text{if }\conj{\mathcal T}{q}\cteq T_j,\\
q-1, & \text{otherwise.}
\end{cases}
\]
Then
\[
\LF{\mathcal T}[i]=\ICA{\mathcal T}[\operatorname{prev}_{\mathcal T}(\CA{\mathcal T}[i])],
\]
and $\FL{\mathcal T}=\LF{\mathcal T}^{-1}$.
The implementation does not store $\LF{\mathcal T}$ explicitly.
Instead, it stores two arrays $\Larr{\mathcal T}$ and $\Farr{\mathcal T}$ whose values are derived from a rotational Cartesian tree signature.
For $X\in(\Sigma\cup\{\$\})^+$, define $\rts{X}[i]$ as
\[
\rts{X}[i]=
\begin{cases}
X[i], & \text{if }X[i]=\$,\\
\rankop_{\infty}(\pde{\rot{X}{i}},|X|)-
\rankop_{\infty}(\pde{X[i]\rot{X}{i}}[2..],|X|), & \text{otherwise.}
\end{cases}
\]
Equivalently, for $X[i]\ne\$$, this is the number of roots on the rightmost-path update that are not smaller than the inserted symbol.
Let $\pi(X)=\rts{X}[1]$.
The arrays stored by the dynamic index are
\[
\Larr{\mathcal T}[i]=\pi\!\left(\conj{\mathcal T}{\CA{\mathcal T}[\LF{\mathcal T}[i]]}\right)
\]
and
\[
\Farr{\mathcal T}[\LF{\mathcal T}[i]]=\Larr{\mathcal T}[i].
\]
They are sufficient to compute $\LF{\mathcal T}$ and $\FL{\mathcal T}$ by rank and select:
\[
\LF{\mathcal T}[i]=\selectop_{\Larr{\mathcal T}[i]}\!\left(\Farr{\mathcal T},
\rankop_{\Larr{\mathcal T}[i]}(\Larr{\mathcal T},i)\right),
\]
\[
\FL{\mathcal T}[i]=\selectop_{\Farr{\mathcal T}[i]}\!\left(\Larr{\mathcal T},
\rankop_{\Farr{\mathcal T}[i]}(\Farr{\mathcal T},i)\right).
\]
This is the Cartesian tree analogue of the LF mapping used by FM-indexes~\cite{ferragina00fmindex,kim21compact,osterkamp26cebwt}.

\cref{tab:rctse-usage} summarizes how the rotational Cartesian tree signature is used in the implemented index, and \cref{tab:rctse-pi-example} gives a concrete computation of $\pi$ for the rotations of $T=5473$.

\begin{table}[t]
    \centering
    \small
    \begin{tabular}{|p{0.32\textwidth}|p{0.58\textwidth}|}
    \hline
    \textbf{Aspect} & \textbf{Purpose in the index} \\
    \hline
    Stored arrays & Store $F$ and $L$ instead of explicit $\FL{\mathcal T}$ and $\LF{\mathcal T}$ arrays. \\
    \hline
    Rotation mapping & Map a conjugate to the corresponding direct rotation through rank and select. \\
    \hline
    Order preservation & Recognize the correct $\omega$-preorder; for example, for $T=4757$, the rotations $(47)$ and $(57)$ are equivalent under $=_{\omega}$. \\
    \hline
    \end{tabular}
    \caption{Role of the rotational Cartesian tree signature encoding in the implemented index. The signature replaces explicit $\FL{\mathcal T}$ and $\LF{\mathcal T}$ arrays by the $F$ and $L$ arrays, maps a conjugate to its previous rotation, and preserves the $\omega$-preorder needed by the circular search space.}
    \label{tab:rctse-usage}
\end{table}

\begin{table}[t]
    \centering
    \small
    \begin{tabular}{|c|c|c|c|c|c|}
    \hline
    i & $T[i]$ & $\rta{T}{i}$ & $\pde{\rta{T}{i}}$ & $\pde{T[i]\cdot\rta{T}{i}}$ & $\rts{T}[i]$ \\
    \hline
    1 & 5 & 4735 & \scalebox{0.50}[1]{$\infty$},1,\scalebox{0.50}[1]{$\infty$},1
    & \scalebox{0.50}[1]{$\infty$},\scalebox{0.50}[1]{$\infty$},1,\scalebox{0.50}[1]{$\infty$},1 & 0 \\
    \hline 
    2 & 4 & 7354 & \scalebox{0.50}[1]{$\infty$},\scalebox{0.50}[1]{$\infty$},1,2 
    & \scalebox{0.50}[1]{$\infty$},1,\scalebox{0.50}[1]{$\infty$},1,2 & 1 \\
    \hline
    3 & 7 & 3547 & \scalebox{0.50}[1]{$\infty$},1,2,1 
    & \scalebox{0.50}[1]{$\infty$},\scalebox{0.50}[1]{$\infty$},1,2,1 & 0 \\
    \hline
    4 & 3 & 5473 & \scalebox{0.50}[1]{$\infty$},\scalebox{0.50}[1]{$\infty$},1,\scalebox{0.50}[1]{$\infty$} 
    & \scalebox{0.50}[1]{$\infty$},1,2,1,4 & 3 \\
    \hline
    \end{tabular}
    \caption{Example computation of the rotational Cartesian tree signature value $\pi$ for the rotations of $T=5473$. The table shows the PDEs of each rotation and of the one-symbol extension used to obtain the signature value.}
    \label{tab:rctse-pi-example}
\end{table}

\subsection{The Dynamic and Static ceBWT Indexes}\label{sec:index-variants}
The dynamic index consists of the arrays
\(\Farr{\mathcal T}, \Larr{\mathcal T},\) and $\LCPinf{\mathcal T}$.
Here $\LCPinf{\mathcal T}[1]=0$, and for $i\ge2$,
\[
\LCPinf{\mathcal T}[i]
=\lcpct{\conj{\mathcal T}{\CA{\mathcal T}[i]}}{\conj{\mathcal T}{\CA{\mathcal T}[i-1]}},
\]
where $\lcpct{X}{Y}$ is the number of $\infty$ symbols in the longest common prefix of $\pde{X}$ and $\pde{Y}$.
This array is used during backward search to resolve intervals when several adjacent conjugates have the same relevant signature prefix.
In the implementation, $\Farr{\mathcal T}$, $\Larr{\mathcal T}$ and $\LCPinf{\mathcal T}$ are stored in dynamic wavelet trees, and an auxiliary dynamic bit vector $E$ is used during incremental construction.

Backward search processes $P$ from right to left.
Assume the current interval is $[\ell..r]=\crange{\mathcal T}{P[i+1..|P|]}$.
For the next symbol, compute
\[
 h=\pi(P[i..|P|]\$),\qquad
 e=\rankop_{\infty}(\pde{P[i..|P|]},|P|-i+1).
\]
The next interval contains exactly the $\LF{\mathcal T}$ images of entries $j\in[\ell..r]$ satisfying
\[
\begin{cases}
\Larr{\mathcal T}[j]=h, & \text{if } e>1,\\
\Larr{\mathcal T}[j]\ge h, & \text{if } e=1.
\end{cases}
\]
The implementation computes the interval boundaries using rank, select, range-count, range-next-value and maximal-interval queries on $\Larr{\mathcal T}$, $\Farr{\mathcal T}$ and $\LCPinf{\mathcal T}$.
The pattern-side computation is illustrated in \cref{tab:pattern-rctse-values}. \cref{tab:example-backward-search} combines these values with the dynamic-index rows of the running example, while \cref{tab:dynamic-search-refinement} shows the interval evolution over successive $\LF{\mathcal T}$ steps.

\begin{table}[t]
    \centering
    \small
    The auxiliary values are $e=\rankop_\infty(i..)$ and $h=\pi(P[i..])$.
    \medskip

    \begin{tabular}{|c|c|c|c|c|c|c|c|}
    \hline
    $i$ & CurrEle[$i$] & PDE & Pre & After & Stack & $e$ & $h/\pi$\\
    \hline
    3 & 1 & \scalebox{0.55}[1]{$\infty$} &  &  & 1 &  1 & 0 \\
    \hline 
    2 & 3 & \scalebox{0.55}[1]{$\infty$},1 & 1 & 1 & 3,1 & 2 & 0 \\
    \hline
    1 & 2 & \scalebox{0.55}[1]{$\infty$},1,\scalebox{0.55}[1]{$\infty$} & 2 & 1 & 2,1 & 2 & 1 \\
    \hline
    \end{tabular}
    \caption{Computation of the pattern-side values used by backward search for $P=231$. The columns show the current element, the PDE, predecessor/successor information maintained by the stack, and the resulting $e$ and $h=\pi(P[i..])$ values.}
    \label{tab:pattern-rctse-values}
\end{table}

\begin{table}[t]
    \centering
    \small
    \textbf{Backward search for pattern 231.}
    
    \par\vspace{0.3cm}
    
    \centering
    \begin{tabular}{|c|c|c|c|c|}
    \hline
    i&P[i..]&$\langle$P[i..]$\rangle$&$e$&h \\
    \hline
    3 & 1 & $\infty$ & 1 & 0\\
    2 & 31 & $\infty,\infty$ & 2 & 0\\
    1 & 231 & $\infty,1,\infty$ & 2 & 1\\
    \hline
    \end{tabular}
    \par\vspace{0.3cm}
    \centering
    \begin{tabular}{|c|c|c|c|c|c|}
    \hline
    $i$ & $\langle\mathrm{CT}[\mathrm{CA}[i]]\rangle$ & $\mathrm{LF}[i]$ & $F[i]$ & $L[i]$ & $\mathrm{LCP}^{\infty}[i]$ \\
    \hline
    1 & \scalebox{0.55}[1]{$\infty$}$,1,2,1$ & 3 & 3 & 0 & 0 \\ 
    \rowcolor{blue!40}
    2 & \scalebox{0.55}[1]{$\infty$}$, 1, $\scalebox{0.6}[1]{$\infty$}$,1$ & 4 & 1 & 0 & 1 \\
    3 & \scalebox{0.55}[1]{$\infty$}$,$\scalebox{0.65}[1]{$\infty$}$,1,2$ & 2 & 0 & 1 & 1 \\
    4 & \scalebox{0.55}[1]{$\infty$}$,$\scalebox{0.65}[1]{$\infty$}$,1,$\scalebox{0.55}[1]{$\infty$} & 1 & 0 & 3 & 2 \\
    \hline
    \end{tabular}
    \caption{Backward-search example for the pattern $231$. The upper table gives the suffix encodings and the derived values $e$ and $h$, while the lower table shows the relevant dynamic-index rows and the selected $\LF{\mathcal T}$ step.}
    \label{tab:example-backward-search}
\end{table}

\begin{table}[t]
    \centering
    \small
    \centering
    \begin{tabular}{|c|c|c|c|c|c|}
    \hline
    $i$ & $\pde{\mathrm{CT}[\mathrm{CA}[i]]}$ & $\mathrm{LF}[i]$ & $F[i]$ & $L[i]$ & $\mathrm{LCP}^{\infty}[i]$ \\
    \hline
    1 & \scalebox{0.55}[1]{$\infty$}$,1,2,1$ & 3 & 3 & 0 & 0 \\ 
    2 & \scalebox{0.55}[1]{$\infty$}$, 1, $\scalebox{0.6}[1]{$\infty$}$,1$ & 4 & 1 & 0 & 1 \\
    \rowcolor{blue!40}
    3 & \scalebox{0.55}[1]{$\infty$}$,$\scalebox{0.65}[1]{$\infty$}$,1,2$ & 2 & 0 & 1 & 1 \\
    \rowcolor{blue!40}
    4 & \scalebox{0.55}[1]{$\infty$}$,$\scalebox{0.65}[1]{$\infty$}$,1,$\scalebox{0.55}[1]{$\infty$} & 1 & 0 & 3 & 2 \\
    \hline
    \end{tabular}
    \par\vspace{0.3cm}
    \begin{tabular}{|c|c|c|c|c|c|}
    \hline
    $i$ & $\pde{\mathrm{CT}[\mathrm{CA}[i]]}$ & $\mathrm{LF}[i]$ & $F[i]$ & $L[i]$ & $\mathrm{LCP}^{\infty}[i]$ \\
    \hline
    1 & \scalebox{0.55}[1]{$\infty$}$,1,2,1$ & 3 & 3 & 0 & 0 \\ 
    \rowcolor{blue!40}
    2 & \scalebox{0.55}[1]{$\infty$}$, 1, $\scalebox{0.6}[1]{$\infty$}$,1$ & 4 & 1 & 0 & 1 \\
    3 & \scalebox{0.55}[1]{$\infty$}$,$\scalebox{0.65}[1]{$\infty$}$,1,2$ & 2 & 0 & 1 & 1 \\
    4 & \scalebox{0.55}[1]{$\infty$}$,$\scalebox{0.65}[1]{$\infty$}$,1,$\scalebox{0.55}[1]{$\infty$} & 1 & 0 & 3 & 2 \\
    \hline
    \end{tabular}
    \caption{Refinement of the dynamic backward-search interval. The upper table shows the interval after applying the $\LF{\mathcal T}$ condition for one suffix, and the lower table shows the narrowed interval after the subsequent step.}
    \label{tab:dynamic-search-refinement}
\end{table}
With dynamic sequence data structures, count queries take
\[
O\!\left(|P|\frac{\lg\sigma\lg n}{\lg\lg n}\right)
\]
time, and the dynamic index occupies $O(n\lg\sigma)$ bits~\cite{osterkamp26cebwt}.

The static index stores the same logical information more compactly.
For an array $A\in\{\Farr{\mathcal T},\Larr{\mathcal T}\}$ and a value $c$, define the bit vector $B_A^c$ over the subsequence of $A$ containing values at least $c$; a bit is $0$ exactly at positions whose value is $c$ and $1$ at positions whose value is larger than $c$.
The families
\(
\{B_{\Farr{\mathcal T}}^c\}_c
\)
and
\(
\{B_{\Larr{\mathcal T}}^c\}_c
\)
encode $\Farr{\mathcal T}$ and $\Larr{\mathcal T}$ in $3n+o(n)$ bits and support the same backward search in $O(|P|)$ time using constant-time rank and select on bit vectors~\cite{kim21compact,osterkamp26cebwt}.
In the implementation, this is the static index represented by the bit-vector collections $B_F$ and $B_L$.
The static representation does not support incremental insertion directly; it is obtained from the dynamic representation and can be converted back only if the additional information needed for reconstruction is retained.
\cref{tab:example-dynamic-static-index} contrasts the dynamic and static representations on the same example.

\begin{table}[t]
    \centering
    \small
    \textbf{Dynamic Index}: 
    \begin{tabular}{|c|c|c|c|c|c|c|}
    \hline
    $i$ & \textbf{$\mathrm{CA}[i]$} & $\mathrm{CT}[\mathrm{CA}[i]]$ & $\langle\mathrm{CT}[\mathrm{CA}[i]]\rangle$ & \textbf{$F[i]$} & \textbf{$L[i]$} & \textbf{$\mathrm{LCP}^{\infty}[i]$} \\
    \hline
    1 & \textbf{4} & 3547 & \scalebox{0.55}[1]{$\infty$}$,1,2,1$ & \textbf{3} & \textbf{0} & \textbf{0} \\ 
    2 & \textbf{2} & 4735 & \scalebox{0.55}[1]{$\infty$}$, 1, $\scalebox{0.6}[1]{$\infty$}$,1$ & \textbf{1} & \textbf{0} & \textbf{1} \\
    3 & \textbf{3} & 7354 & \scalebox{0.55}[1]{$\infty$}$,$\scalebox{0.65}[1]{$\infty$}$,1,2$ & \textbf{0} & \textbf{1} & \textbf{1}\\
    4 & \textbf{1} & 5473 & \scalebox{0.55}[1]{$\infty$}$,$\scalebox{0.65}[1]{$\infty$}$,1,$\scalebox{0.55}[1]{$\infty$} & \textbf{0} & \textbf{3} & \textbf{2}\\
    \hline
    \end{tabular}

    \vspace{1em}
    \textbf{Static Index}:\\
    \centering
    \par\vspace{0.3cm}
    \centering
    \begin{minipage}{0.45\textwidth}
    \centering
    \begin{tabular}{|c|cccc|}
    \hline
     & 1 & 2 & 3 & 4 \\
    \hline
    $F$ & 3 & 1 & 0 & 0 \\
    \hline
    $B_F$ ($c=0$) & \one & \one & \zero & \zero \\
    $B_F$ ($c=1$) & \one & \zero & \emptyc & \emptyc \\
    $B_F$ ($c=2$) & \one & \emptyc & \emptyc & \emptyc \\
    $B_F$ ($c=3$) & \zero & \emptyc & \emptyc & \emptyc \\
    \hline
    \end{tabular}
    \end{minipage}
    \vspace{1cm}
    \begin{minipage}{0.45\textwidth}
    \centering
    \begin{tabular}{|c|cccc|}
    \hline
     & 1 & 2 & 3 & 4 \\
    \hline
    $L$ & 0 & 0 & 1 & 3 \\
    \hline
    $B_L$ ($c=0$) & \zero & \zero & \one & \one \\
    $B_L$ ($c=1$) & \emptyc & \emptyc & \zero & \one \\
    $B_L$ ($c=2$) & \emptyc & \emptyc & \emptyc & \one\\
    $B_L$ ($c=3$) & \emptyc & \emptyc & \emptyc & \zero \\
    \hline
    \end{tabular}
    \end{minipage}
    \caption{Worked example of the dynamic and static ceBWT representations for the circular text rotations of $3547$, $4735$, $7354$ and $5473$. The dynamic part lists the conjugate array, encoded rotations, $F$, $L$, and $\mathrm{LCP}^{\infty}$ values. 
        The static ceBWT consists of the $B_F^c$ and $B_L^c$ rows encoding the $F$ and $L$ arrays and supports constant-time rank and select during static backward search.
    }
    \label{tab:example-dynamic-static-index}
\end{table}

\subsection{Construction Model Used by the Implementation}\label{sec:construction-model}
For a single text $T$ of length $n$, the theoretical construction builds an auxiliary text $R=T^4\$$.
The fourfold copy guarantees enough symbols to compare conjugates under the $\omega$-preorder, since equality of two conjugates of length at most $n$ is decided by the first $3n$ encoded symbols.
After constructing the index of $R$, only the entries corresponding to the true conjugates of $T$ are retained.
For multiple texts, the construction inserts texts incrementally, merging the newly constructed single-text index into the existing index while updating $\Farr{\mathcal T}$, $\Larr{\mathcal T}$, $\LCPinf{\mathcal T}$ and the auxiliary insertion bit vector $E$.
The theoretical construction time is
\[
O\!\left(n\frac{\lg\sigma\lg n}{\lg\lg n}\right)
\]
using $O(n\lg\sigma)$ bits of working space.
The correctness reference for the implementation is this fourfold construction. The experiments additionally evaluate engineering variants: 
\begin{itemize}
    \item a fourfold construction based on \cite{osterkamp26cebwt}, 
    \item a threefold construction and \twoFold{} supported by our theoretical results in \cref{sec:folding}, and
    \item a threaded variant of \twoFold{}, in which partial indexes are built in parallel and merged sequentially. 
\end{itemize}
The reduced variants are implementation optimizations; they do not change the logical index described above.
For the reduced construction, \cref{tab:folding-input-output} records the inputs and outputs of the folding computations, \cref{tab:folding-rctse-computation} shows the signature computation. The insertion process is expanded in \cref{tab:insertion-base-index,tab:insert-single-index,tab:insert-merged-index,tab:insert-folding-view,tab:insert-comparison-existing,tab:insert-comparison-new}.

\begin{table}[t]
    \centering
    \small
    \begin{tabular}{|p{0.23\textwidth}|p{0.33\textwidth}|p{0.34\textwidth}|}
    \hline
    \textbf{Computation} & \textbf{Input} & \textbf{Output} \\
    \hline
    Index entry for each symbol & $\pi(bR)$, $CA^{-1}$, $F$, $L$, $\mathrm{LCP}^{\infty}$ & $F_{\mathrm{new}}$, $L_{\mathrm{new}}$, $\mathrm{LCP}^{\infty}_{\mathrm{new}}$, and the corresponding $CA$ value \\
    \hline
    Comparison values for each symbol & $\pi(R)$, $\cntarr{i}$, $\plcparr{i}$, $\slcparr{i}$, $F_{\mathrm{old}}$, $L_{\mathrm{old}}$, $\mathrm{LCP}^{\infty}_{\mathrm{old}}$ & $\cntarr{i+1}$, $\plcparr{i+1}$, and $\slcparr{i+1}$ \\
    \hline
    \end{tabular}
    \caption{Inputs and outputs used by the folding constructor. The index update computes new $F$, $L$, $\mathrm{LCP}^{\infty}$, and conjugate-array values, while the comparison-value update maintains the count and special LCP values needed during merging.}
    \label{tab:folding-input-output}
\end{table}

\begin{table}[t]
    \centering
    \small
    \centering
    \begin{tabular}{|c|c|c|c|c|c|}
    \hline
    $i$ & CurrEle & Pre & After & Stack & $\pi$ \\
    \hline
    8 & 3 & 1 & 0 & 3 &  \\
    \hline 
    7 & 7 & 1 & 1 & 7,3 & 0 \\
    \hline
    6 & 4 & 2 & 1 & 4,3 & 1 \\
    \hline
    5 & 5 & 2 & 2 & 5,4,3 & 0 \\
    \hline
    4 & 3 & 3 & 0 & 3 & 3 \\
    \hline
    3 & 7 & 1 & 1 & 7,3 & 0 \\
    \hline
    2 & 4 & 2 & 1 & 4,3 & 1 \\
    \hline
    1 & 5 & 2 & 2 & 5,4,3 & 0 \\
    \hline
    \end{tabular}
    \caption{Right-to-left computation of rotational Cartesian tree signature values. The stack records the maintained rightmost path, and the $PI$ column gives the resulting signature value for each processed position.}
    \label{tab:folding-rctse-computation}
\end{table}

\begin{table}[t]
    \centering
    \small
    \begin{tabular}{|c|c|c|c|c|c|c|}
    \hline
    $i$ & $\mathrm{CT}[\mathrm{CA}[i]]$ & $\pde{\mathrm{CT}[\mathrm{CA}[i]]}$ & $\mathrm{FL}[i]$ & $F[i]$ & $L[i]$ & $\mathrm{LCP}^{\infty}[i]$ \\
    \hline
    1 & 5586 & \scalebox{0.50}[1]{$\infty$}$,1,1,2$ & 4 & 1 & 0 & 0\\
    2 & 3734 & \scalebox{0.50}[1]{$\infty$}$,1,2,1$ & 6 & 2 & 0 & 1\\
    3 & 3437 & \scalebox{0.50}[1]{$\infty$}$,1,2,1$ & 7 & 2 & 0 & 1\\
    4 & 5865 & \scalebox{0.50}[1]{$\infty$}$,1,2,3$ & 8 & 3 & 1 & 1\\
    5 & 6558 & \scalebox{0.55}[1]{$\infty$}$,$\scalebox{0.55}[1]{$\infty$}$,1,1$ & 1 & 0 & 0 & 1\\
    6 & 4373 & \scalebox{0.55}[1]{$\infty$}$,$\scalebox{0.55}[1]{$\infty$}$,1,2$ & 2 & 0 & 2 & 2\\
    7 & 7343 & \scalebox{0.55}[1]{$\infty$}$,$\scalebox{0.55}[1]{$\infty$}$,1,2$ & 3 & 0 & 2 & 2\\
    8 & 8655 & \scalebox{0.55}[1]{$\infty$}$,$\scalebox{0.55}[1]{$\infty$}$,$\scalebox{0.55}[1]{$\infty$}$,1$ & 5 & 0 & 3 & 2\\
    \hline
    \end{tabular}
    \caption{Base dynamic index before insertion in the worked example. The rows list the current encoded rotations together with their $\mathrm{FL}$, $F$, $L$, and $\mathrm{LCP}^{\infty}$ values.}
    \label{tab:insertion-base-index}
\end{table}

\begin{table}[t]
    \centering
    \begin{framed}
        \begin{itemize}
            \item add \textbf{547}
            \item build the index for \textbf{547}
        \end{itemize}
        \centering
        \begin{tabular}{|c|c|c|c|c|c|c|}
        \hline
        $i$ & $\mathrm{CA}[i]$ & $\mathrm{CT}[\mathrm{CA}[i]]$ & $\pde{\mathrm{CT}[\mathrm{CA}[i]]}$ & $F[i]$ & $L[i]$ & $\mathrm{LCP}^{\infty}[i]$ \\
        \hline
        1 & 2 & 475 & \scalebox{0.50}[1]{$\infty$}$,1,2$ & 3 & 0 & 0\\
        2 & 1 & 547 & \scalebox{0.55}[1]{$\infty$}$,$\scalebox{0.55}[1]{$\infty$}$,1$ & 0 & 0 & 1\\
        3 & 3 & 754 & \scalebox{0.55}[1]{$\infty$}$,$\scalebox{0.55}[1]{$\infty$}$,$\scalebox{0.55}[1]{$\infty$} & 0 & 3 & 2\\
        \hline
        \end{tabular}
        
        \par\vspace{0.1cm}
        
        \begin{itemize}
            \item compute $\cntarr{i}$ and special LCP values
        \end{itemize}
        \begin{tabular}{|c|c|c|c|c|c|}
        \hline
        $i$ & $\mathrm{CA}[i]$ && $\cntarr{i}$ & $\plcparr{\mathrm{CA}[i]}$ & $\slcparr{\mathrm{CA}[i]}$ \\
        \hline
        1 & 2 && 4 & 1 & 1 \\
        2 & 1 && 7 & 2 & 2 \\
        3 & 3 && 8 & 3 & -1 \\
        \hline
        \end{tabular}
    \end{framed}
    \caption{Single-text index built for the sequence $547$ before it is merged into the existing dynamic index. The lower table contains the count and neighboring LCP values used by the merge step.}
    \label{tab:insert-single-index}
\end{table}

\begin{table}[t]
    \centering
    \small
    \begin{tabular}{|c|c|c|c|c|c|c|}
    \hline
    $i$ & $\mathrm{CT}[\mathrm{CA}[i]]$ & $\pde{\mathrm{CT}[\mathrm{CA}[i]]}$ & $\mathrm{FL}[i]$ & $F[i]$ & $L[i]$ & $\mathrm{LCP}^{\infty}[i]$ \\
    \hline
    1 & 5586 & \scalebox{0.50}[1]{$\infty$}$,1,1,2$ & 4 & 1 & 0 & 0\\
    2 & 3734 & \scalebox{0.50}[1]{$\infty$}$,1,2,1$ & 7 & 2 & 0 & 1\\
    3 & 3437 & \scalebox{0.50}[1]{$\infty$}$,1,2,1$ & 8 & 2 & 0 & 1\\
    4 & 5865 & \scalebox{0.50}[1]{$\infty$}$,1,2,3$ & 10 & 3 & 1 & 1\\
    \rowcolor{blue!40}
    5 & 475 & \scalebox{0.50}[1]{$\infty$}$,1,2,3$ & 11 & 3 & 0 & 1 \\
    6 & 6558 & \scalebox{0.55}[1]{$\infty$}$,$\scalebox{0.55}[1]{$\infty$}$,1,1$ & 1 & 0 & 0 & 1\\
    7 & 4373 & \scalebox{0.55}[1]{$\infty$}$,$\scalebox{0.55}[1]{$\infty$}$,1,2$ & 2 & 0 & 2 & 2\\
    8 & 7343 & \scalebox{0.55}[1]{$\infty$}$,$\scalebox{0.55}[1]{$\infty$}$,1,2$ & 3 & 0 & 2 & 2\\
    \rowcolor{blue!40}
    9 & 547 & \scalebox{0.55}[1]{$\infty$}$,$\scalebox{0.55}[1]{$\infty$}$,1,2$ & 5 & 0 & 0 & 2 \\
    10 & 8655 & \scalebox{0.55}[1]{$\infty$}$,$\scalebox{0.55}[1]{$\infty$}$,$\scalebox{0.55}[1]{$\infty$}$,1$ & 6 & 0 & 3 & 2\\
    \rowcolor{blue!40}
    11 & 754 & \scalebox{0.55}[1]{$\infty$}$,$\scalebox{0.55}[1]{$\infty$}$,$\scalebox{0.55}[1]{$\infty$}$,1$ & 9 & 0 & 3 & 3\\
    \hline
    \end{tabular}
    \caption{Merged dynamic index after inserting the sequence $547$. Highlighted rows correspond to the rotations contributed by the inserted sequence.}
    \label{tab:insert-merged-index}
\end{table}

\begin{table}[t]
    \centering
    \small
    \centering
    \begin{tabular}{|c|c|c|c|c|c|}
    \hline
    $i$ & $\mathrm{CT}[i]$ & $\mathrm{CA}[i]$ & $F[i]$ & $L[i]$ & $\mathrm{LCP}^{\infty}[i]$ \\
    \hline
    1 & 547547\$ & 7 & \$ & 0 & 0\\
    2 & 47547\$5 & 6 & 0 & 1 & 0\\
    3 & 7547\$54 & 5 & 1 & 0 & 1\\
    \rowcolor{blue!40}
    4 & 547\$547 & 2 & 3 & 0 & 1\\
    \rowcolor{blue!40}
    5 & 47\$5475 & 4 & 0 & 0 & 1\\
    6 & 7\$54754 & 1 & 0 & \$ & 2\\
    \rowcolor{blue!40}
    7 & \$547547 & 3 & 0 & 3 & 2\\
    \hline
    \end{tabular}
    \caption{Folding view of the insertion example. The table shows the replicated text used during construction and the retained rows that become the final entries of the inserted sequence.}
    \label{tab:insert-folding-view}
\end{table}

\begin{table}[t]
    \centering
    \begin{tabular}{|c|c|c|c|c|}
    \hline
    $i$ & $\rta{CT}{i}$ & $\cntarr{i}$ & $\plcparr{i}$ & $\slcparr{i}$ \\
    \hline
    9 & \$547547547 & 0 & -1 & 0\\
    8 & 7\$54754754 & 0 & -1 & 0\\
    7 & 47\$5475475 & 0 & -1 & 0\\
    6 & 547\$547547 & 4 & 1 & 2\\
    5 & 7547\$54754 & 7 & 2 & 3\\
    4 & 47547\$5475 & 3 & 1 & 1\\
    \rowcolor{blue!40}
    3 & 547547\$547 & 7 & 2 & 2\\
    \rowcolor{blue!40}
    2 & 7547547\$54 & 8 & 3 & -1\\
    \rowcolor{blue!40}
    1 & 47547547\$5 & 4 & 1 & 1\\
    \hline
    \end{tabular}
    \caption{Comparison values computed against the existing index during insertion. The highlighted rows are the values associated with the new sequence before extraction and merging.}
    \label{tab:insert-comparison-existing}
\end{table}

\begin{table}[t]
    \centering
    \begin{tabular}{|c|c|c|c|c|c|}
    \hline
    $i$ & $\rta{CT}{i}$ & $\mathrm{CA}[i]$ & $\cntarr{i}$ & $\plcparr{i}$ & $\slcparr{i}$ \\
    \hline
    1 & 475 & 2 & 4 & 1 & 1\\
    2 & 754 & 1 & 8 & 3 & -1\\
    3 & 547 & 3 & 7 & 2 & 2\\
    \hline
    \end{tabular}
    \par\vspace{0.3cm}
    \begin{tabular}{|c|c|c|c|c|c|}
    \hline
    $i$ & $\rta{CT}{i}$ & $\mathrm{CA}[i]$ & $\cntarr{i}$ & $\plcparr{i}$ & $\slcparr{i}$ \\
    \hline
    1 & 547 & 2 & 4 & 2 & 2\\
    2 & 475 & 1 & 7 & 1 & 1\\
    3 & 754 & 3 & 8 & 3 & -1\\
    \hline
    \end{tabular}
    \caption{Comparison values for the rotations of the inserted sequence, shown before and after the cyclic shift used by the merge computation.}
    \label{tab:insert-comparison-new}
\end{table}

When a new text is merged into an existing index, we use the following helper values from the theoretical construction. For a string $X=x_1x_2\cdots x_m$, the notation
\[
\rta{X}{i}=x_{i+1}x_{i+2}\cdots x_mx_1\cdots x_i
\]
denotes the $i$-th left rotation of $X$, with indices understood modulo $m$; this is the same conjugate as $\rot{X}{i}$ above. Let $\mathcal R$ be the already indexed collection and let $\rho=|\CA{\mathcal R}|$ be its total number of conjugates. For any comparison string $V$, define
\[
\cnt{\mathcal R}{V}=\left|\left\{j\in[1..\rho]\mid
\conj{\mathcal R}{\CA{\mathcal R}[j]}\ctpeq V\right\}\right|.
\]
Thus $\cnt{\mathcal R}{V}$ is the insertion rank of $V$ among the old conjugates. The predecessor and successor LCP helper values are
\[
\begin{aligned}
\plcp{\mathcal R}{V}
&=
\begin{cases}
-1, & \text{if }\cnt{\mathcal R}{V}=0,\\
\lcpct{\conj{\mathcal R}{\CA{\mathcal R}[\cnt{\mathcal R}{V}]}}{V}, & \text{otherwise,}
\end{cases}\\[0.4em]
\slcp{\mathcal R}{V}
&=
\begin{cases}
\lcpct{\conj{\mathcal R}{\CA{\mathcal R}[\cnt{\mathcal R}{V}+1]}}{V}, & \text{if }\cnt{\mathcal R}{V}<\rho,\\
-1, & \text{otherwise.}
\end{cases}
\end{aligned}
\]
Here $\lcpct{U}{W}$ counts the number of $\infty$ symbols in the common prefix of the PDEs of $U$ and $W$. In example tables, entries such as $\cntarr{i}$, $\plcparr{i}$, and $\slcparr{i}$ abbreviate these three values for the corresponding rotation $\rta{S}{i}$ of the text $S$ being inserted.

\section{Implementing the ceBWT-index}
\label{section:softwarearchitecture}

We present the ceBWT in two main versions: as a dynamic index and as a static index. 
We can convert the dynamic index into the static, and vice versa.
Each of the two indexes is further given with three types of storage variants, 
which differ in the amount of auxiliary information they retain and thus in the operations they can support with exact or weakened semantics.
We will later see in \cref{sec:storage} that we can rebuild this auxiliary information up to CTM-equivalence, rotation and exponentiation from the root of the encoded data.
In \cref{sec:folding}, we further optimize the construction of the dynamic index by temporarily indexing less information,
and propose a parallel construction in \cref{sec:parallelism}.

\subsection{Building Blocks}
\label{section:implementationdifferences}

The static and dynamic ceBWT index share a storage module.
In the storage module, we define three storage variants: \texttt{FullStorage}, \texttt{CAStorage} and \texttt{MinimalStorage}. 
A storage variant determines which auxiliary information is retained in addition to the core \textsc{ceBWT} structures and, consequently, which operations can be answered with exact or weakened semantics; see \cref{tab:storage}.

Let $\textsf{Occ}(P)$ denote the set of all occurrences of a pattern $P$ in the circular search space of the indexed texts. An occurrence is represented by a pair $(j,p)$, where $j$ is a text identifier and $p$ is a starting position in $T_j$. If a match crosses the end of $T_j$, the reported position is still $p$ and the corresponding substring is interpreted circularly as the length-$|P|$ prefix of $T_j^\omega[p..]$.
We use primes in operation names to indicate weaker guarantees: an unprimed operation is exact with respect to the original indexed strings, a singly primed operation is exact only up to Cartesian-tree matching equivalence, and a doubly primed operation guarantees only that some indexed string containing a circular occurrence is affected.

The supported operations are as follows:
\begin{itemize}
    \item $\operatorname{count}(P)$ returns $|\textsf{Occ}(P)|$.
    \item $\operatorname{locate}(P)$ returns the pairs in $\textsf{Occ}(P)$.
    \item $\operatorname{extract}(P)$ returns, for each $(j,p)\in\textsf{Occ}(P)$, the length-$|P|$ substring of $T_j^\omega$ starting at $p$.
    \item $\operatorname{decode}(i)$ returns the original indexed string $T_i$.
    \item $\operatorname{decode}'(i)$ returns a representative string $\widehat T_i$ with $\widehat T_i\ctmatch T_i$. Thus, the Cartesian-tree shape is preserved, but the original alphabet values are not guaranteed to be recovered.
    \item $\operatorname{remove}(P)$ removes an indexed string that is exactly equal to $P$.
    \item $\operatorname{remove}'(P)$ removes an indexed string $T_i$ with $T_i\ctmatch P$; if several indexed strings satisfy this condition, the choice is implementation-dependent.
    \item $\operatorname{remove}''(P)$ removes some indexed string $T_i$ for which $\textsf{Occ}(P)$ contains an occurrence in $T_i$. This operation does not require $|T_i|=|P|$ and does not guarantee that $T_i\ctmatch P$ as a whole.
\end{itemize}
See \cref{figExRemoval} for an example of the different removal operations.

Full storage stores the input strings, the conjugate array and the sequence endpoints. It therefore supports exact counting, locating, extraction, decoding and removal. CA storage stores the conjugate array and endpoints, but not the input strings. It can still locate occurrences and identify text boundaries, but decoding and removal are only possible up to Cartesian-tree matching equivalence. The minimal storage keeps only the information required by the core index. It supports counting and matching-based deletion, but it cannot report exact occurrence positions, recover original alphabet values, or identify the $i$-th input string well enough to support $\operatorname{decode}'(i)$.

\begin{table}[t]
    \centering
    \footnotesize
    \begin{tabular}{|p{2.7cm}|p{3.6cm}|p{3.6cm}|p{3.6cm}|}
        \hline
        \textbf{Storage variant} & \textbf{Full storage} & \textbf{CA storage} & \textbf{Minimal storage} \\
        \hline
        \textbf{Stored information}
        & Input strings; conjugate array; endpoints
        & Conjugate array; endpoints
        & Core index only \\
        \hline
        \textbf{Operations}
        & $\operatorname{count}(P)$\newline
          $\operatorname{locate}(P)$\newline
          $\operatorname{extract}(P)$\newline
          $\operatorname{decode}(i)$\newline
          $\operatorname{remove}(P)$
        & $\operatorname{count}(P)$\newline
          $\operatorname{locate}(P)$\newline
          $\operatorname{decode}'(i)$\newline
          $\operatorname{remove}'(P)$
        & $\operatorname{count}(P)$\newline
          $\operatorname{remove}''(P)$ \\
        \hline
        \textbf{Guarantee}
        & Exact with respect to the indexed strings
        & Exact up to Cartesian-tree matching equivalence
        & Matching-based deletion only; no locations, string identifiers, or original values \\
        \hline
    \end{tabular}
    \caption{Supported operations of the three storage variants.}
    \label{tab:storage}
\end{table}

\begin{figure}[t]
    \centering
    \begin{tabular}{|p{2.0cm}|p{2cm}p{2cm}p{6cm}|}
        \hline
        operation / type & full & CA & minimal \\
        \hline
        delete $P$ & $\operatorname{remove}(P)$ & $\operatorname{remove}'(P)$ & $\operatorname{remove}''(P)$ \\
        \hline
        delete $S_1$ & $S_1$ & $S_1$ or $S_2$ & any text containing a match \\
        \hline
        delete $S_3$ & $S_3$ & $S_3$ & any text containing a match \\
        \hline
    \end{tabular}
    \caption{Examples of the remove operation.
    Assume that the index contains the following three strings 
    $S_1 = [1, 2, 1, 2]$,
    $S_2 = [2, 3, 2, 3]$, and
    $S_3 = [1, 2]$.
    The second row specifies the function called when deleting a pattern $P$ from each storage variant.
    The other rows show, for each such deletion operation, which indexed string(s) are subject to removal.
    The operation $\operatorname{remove}(P)$ for full storage removes only a string exactly equal to $P$, whereas
    $\operatorname{remove}'(S_1)$ may remove any string that matches $S_1$ up to Cartesian-tree matching equivalence, e.g., $S_1 \ctmatch S_2$.
    The operation $\operatorname{remove}''(S_3)$ is weaker: since $S_3$ occurs as a circular Cartesian-tree match inside $S_1$ and $S_2$, it may remove one of these longer strings instead of the indexed string $S_3$.
    }
    \label{figExRemoval}
\end{figure}

In the utility module, we provide general-purpose functions, a \texttt{BitVectorSupport} structure for rank and select operations on static bit vectors implemented with SDSL~\cite{gog14sdsl}, and a \texttt{Mapper} for transforming genomic input sequences or integer values in string format into integer types and back.

In the core module, we implement templates for evaluating pattern search and managing the data structures of both indices, as well as the extra classes for removing and locating sequences. These functions depend strongly on the storage variant, because it determines the achievable level of precision.

The sequences and conjugate arrays are stored in vectors. We store the sequences as copies of the input, while the conjugate array is stored as integer values. An additional vector records the positions of each newly added sequence so that they do not need to be recomputed by the functions implemented in the core module. 

The dynamic data structure is primarily represented by three dynamic wavelet trees and is supplemented with a dynamic bit vector. This structure corresponds to $L$, $F$, $\mathrm{LCP}^{\infty}$, and $e$. 
The $L$ and $F$ structures are used to traverse the index by LF mapping over all conjugates of a sequence, as described in~\cite[Section 3.2]{osterkamp26cebwt}. The $\mathrm{LCP}^{\infty}$ array is used during backward search to resolve ambiguities when determining the range of matching conjugates, because it stores the longest common prefix length between adjacent conjugates.
While all three are mandatory for the dynamic index, $e$ serves as a helper structure. It is a zeroed bit vector used when adding a new sequence to efficiently determine the correct positions for inserting the new sequence's information into the wavelet trees. 
Creating this object on demand introduced some overhead, so we store it persistently within the dynamic index structure.

\begin{figure}[htbp]
    \centering
    \fbox{\begin{tikzpicture}

\draw[rounded corners, fill=cyan!40] (0, 3) rectangle (2.5, 5);
    \node[align=center] at (1.25, 4) {Dynamic \\ ceBWT};
    
\draw[fill=cyan!20] (0, 0.1) rectangle (0.55, 2.9);
    \node[rotate=-90, font=\small] at (0.27, 1.5) {$F$};
    \draw[fill=cyan!20] (0.6, 0.1) rectangle (1.15, 2.9);
    \node[rotate=-90, font=\small] at (0.87, 1.5) {$L$};
    \draw[fill=cyan!20] (1.2, 0.1) rectangle (1.75, 2.9);
    \node[rotate=-90, font=\small] at (1.47, 1.5) {$\mathrm{LCP}^{\infty}$};
    
\draw[fill=green!30] (1.8, 0.1) rectangle (2.35, 2.9);
    \node[rotate=-90, font=\small] at (2.07, 1.5) {$e$};
    
\draw[rounded corners, fill=red!40] (3.5, 3) rectangle (6.5, 5);
    \node[align=center] at (5, 4) {Static \\ ceBWT};
    
\draw[fill=orange!40] (3.5, 0.1) rectangle (4.05, 2.9);
    \node[rotate=-90, font=\small] at (3.77, 1.5) {$B_F$};
    \draw[fill=orange!40] (4.1, 0.1) rectangle (4.65, 2.9);
    \node[rotate=-90, font=\small] at (4.37, 1.5) {$B_L$};
    
\draw[fill=cyan!20] (5.3, 0.1) rectangle (5.85, 2.9);
    \node[rotate=-90, font=\small] at (5.57, 1.5) {$\mathrm{LCP}^{\infty}$};
    
\draw (7, 0.5) rectangle (12, 3);
    \node[font=\small] at (8.75, 2.7) {\textbf{Legend}};
    
    \draw[fill=cyan!20] (7.2, 2.1) rectangle (7.6, 2.4);
    \node[anchor=west, font=\small] at (7.7, 2.25) {Dynamic Wavelet Tree};
    
    \draw[fill=orange!40] (7.2, 1.5) rectangle (7.6, 1.8);
    \node[anchor=west, font=\small] at (7.7, 1.65) {Static Bitvector Collection};
    
    \draw[fill=green!30] (7.2, 0.9) rectangle (7.6, 1.2);
    \node[anchor=west, font=\small] at (7.7, 1.05) {Dynamic Bitvector};
    
    \end{tikzpicture}
    }
    \caption{Dynamic and static ceBWT data structures used in the implementation.}
    \label{fig:dataStructureimpl}
\end{figure}
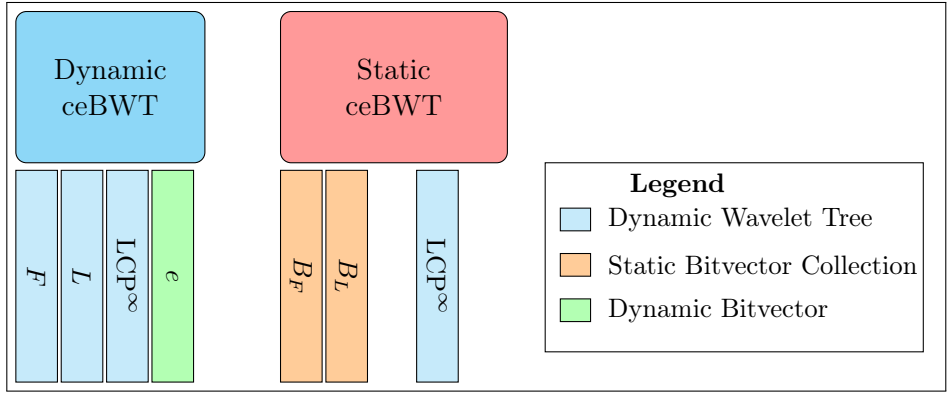

The static data structure is built according to the construction described in~\cite[Section 5]{kim21compact}, replacing the dynamic wavelet trees with static bit vector collections $B_F$ and $B_L$. These bit vectors encode the $L$ and $F$ arrays in a space-efficient unary representation, enabling efficient rank and select queries and supporting the same backward search as the dynamic index. To allow switching back to the dynamic index, the previously generated $\mathrm{LCP}^{\infty}$ wavelet tree must be temporarily stored. This is necessary because generating $\mathrm{LCP}^{\infty}$ requires more information than is available from the stored $B_F$ and $B_L$ bit vectors. We do not convert it into a static wavelet tree, as this would require another transformation when switching back to the dynamic index, and the runtime overhead would outweigh the memory saving (see \cref{fig:dataStructureimpl}).

\subsection{Reconstructing the Index from $\Larr{\mathcal T}$}\label{sec:storage}
For completely freezing the index onto disk for long-term storage, we propose to keep
only the $\Larr{\mathcal T}$ array, from which we can reconstruct the full index up to CTM-equivalence and exponentiation of the indexed collection $\mathcal T$. 
The crucial part is the reconstruction of $\Farr{\mathcal T}$, from which we regain the LF mapping for the static index, and from there we can reconstruct the dynamic index.
We claim the following.

\begin{lemma}
    Given $\Larr{\mathcal T}$, we can reconstruct $\Farr{\mathcal T}$ in $O(n^2)$ time, where $n$ is the total length of the indexed collection $\mathcal T$.
\end{lemma}

For simplicity, assume that the indexed collection $\mathcal T=\{T_1,\ldots,T_d\}$ with $T_k\in\Sigma^+$ satisfies the following two properties:
\begin{itemize}
	\item $n_k=  \left|\primitive{\rpde{T_k}}\right|$ for each $1\leq k\leq d$.
	\item $T_i\not\cteq \rta{T_j}{x}$ for all $1\leq i,j\leq d$ and $x\in\mathbb{N}_0$ with $i\neq j$.
\end{itemize}
The first assumption removes repetitions inside one text, and the second assumption excludes two distinct indexed texts from being equivalent up to CTM-equivalence and rotation.
Under these assumptions, we can reconstruct the missing static structures from $\Larr{\mathcal T}$ by adapting the reconstruction technique of Kawanami et al.~\cite{kawanami25inverting}.

Let $z=\max\{n_k\mid 1\leq k\leq d\}$ and $n=\sum_{k=1}^{d}n_k$.
For $x\in\mathbb{N}_0$, define the multiset
\[
\begin{split}
	\mathcal P_x=\{&\pde{\rta{T_1}{1}^{\omega}[1..x]},\ldots,\pde{\rta{T_1}{n_1}^{\omega}[1..x]},\\
	&\pde{\rta{T_2}{1}^{\omega}[1..x]},\ldots,\pde{\rta{T_2}{n_2}^{\omega}[1..x]},\\
	&\ldots,\\
	&\pde{\rta{T_d}{1}^{\omega}[1..x]},\ldots,\pde{\rta{T_d}{n_d}^{\omega}[1..x]}\}.
\end{split}
\]
For every $x\in\mathbb{N}_0$ and $1\leq i\leq n$, the $i$-th smallest element of $\mathcal P_x$ is
\[
	\pde{\conj{\mathcal T}{\CA{\mathcal T}[i]}^{\omega}[1..x]}.
\]
Consequently, $\mathcal P_{x+1}$ can be obtained from $\mathcal P_x$ and $\Larr{\mathcal T}$ as follows.
Consider the $i$-th smallest element
\[
	\pde{\conj{\mathcal T}{\CA{\mathcal T}[i]}^{\omega}[1..x]}
\]
of $\mathcal P_x$.
Let $r_i=\rankop_{\infty}(\pde{\conj{\mathcal T}{\CA{\mathcal T}[i]}^{\omega}[1..x]},x)$.
Then:
\begin{itemize}
	\item If $\Larr{\mathcal T}[i]\geq r_i$, then
	\[
		\pde{\conj{\mathcal T}{\CA{\mathcal T}[\LF{\mathcal T}[i]]}^{\omega}[1..x+1]}
		=\infty\cdot W,
	\]
	where $W$ is obtained from $\pde{\conj{\mathcal T}{\CA{\mathcal T}[i]}^{\omega}[1..x]}$ by replacing each occurrence $j$ of $\infty$ by $j$.
	\item If $\Larr{\mathcal T}[i]< r_i$, then
	\[
		\pde{\conj{\mathcal T}{\CA{\mathcal T}[\LF{\mathcal T}[i]]}^{\omega}[1..x+1]}
		=\infty\cdot W,
	\]
	where $W$ is obtained from $\pde{\conj{\mathcal T}{\CA{\mathcal T}[i]}^{\omega}[1..x]}$ by replacing the first $\Larr{\mathcal T}[i]$ occurrences $j$ of $\infty$ by $j$.
\end{itemize}

By the assumptions above and by~\cite[Lemma 3.3]{osterkamp26cebwt}, the multiset $\mathcal P_x$ is in fact a set of $n$ elements for every $x\geq 3z$.
Moreover, the same reconstruction step allows us to infer the LF mapping from $\mathcal P_{3z}$, and the LF mapping can then be used to compute $\Farr{\mathcal T}$.
Thus, $\Farr{\mathcal T}$ can be reconstructed from $\Larr{\mathcal T}$ in $O(n^3)$ time and $O(n^2)$ bits of space.
Using an implicit representation of $\mathcal P_x$, the time can be improved to $O(n^2)$ and the working space to $O(n)$ bits, as in Theorem~6 of Kawanami et al.~\cite{kawanami25inverting}.

Finally, to reconstruct the dynamic index, 
we use the LF mapping to recover the indexed texts up to CTM-equivalence, rotation and text order, and then rebuild the dynamic index for the resulting collection.
Similarly, we can rebuild $\mathrm{CA}$ up to CTM-equivalence and exponentiation of the indexed texts (cf.\ the example for the remove operation in \cref{figExRemoval})
by using the LF mapping to recover all indexed texts up to CTM-equivalence and rotation, and writing decremental numbers (with modulo) in $\mathrm{CA}$ for the starting position of each extracted conjugate.
Similarly, we can rebuild $\mathrm{LCP}^{\infty}$ by extracting and processing each input text individually.
This all can be done in $O(n)$ time.
To actually regain the original $\mathrm{CA}$ (or original $\mathrm{LCP}^{\infty}$) in general, we need additional information on the indexed texts, 
such as the length of each indexed text and the number of occurrences of each indexed text in $\mathrm{CA}$.

\subsection{Foldings}\label{sec:folding}

\begin{figure}[t]

    \begin{minipage}{0.48\linewidth}
    \begin{tabular}{|c|c|c|c|c|c|} 
    \hline 
    $i$ & $\mathrm{CA}[i]$ && $F[i]$ & $L[i]$ & $\mathrm{LCP}^{\infty}[i]$ \\ 
    \hline 
    1 & 11 && \$ & 0 & 0 \\ 2 & 10 && 0 & 0 & 0 \\ 
    \rowcolor{blue!40}
    3 & 6 && \tikzmark{startF}3 & 0 & 1 \\ 
    4 & 1 && 4 & \$ & 1 \\ 
    5 & 8 && 1 & 0 & 1 \\ 
    \rowcolor{blue!40} 
    6 & 3 && 1 & 0 & 2 \\ 
    7 & 9 && 0 & 1 & 1 \\ 
    \rowcolor{blue!40} 
    8 & 5 && 0 & 0 & 2 \\ 
    9 & 7 && 0 & 3 & 2 \\ 
    \rowcolor{blue!40} 10 & 2 && 0 & \tikzmark{endL}4 & 3 \\ 
    \rowcolor{blue!40} 11 & 4 && 0 & 1 & 2 \\ 
    \hline 
    \end{tabular} 
    
    \begin{tikzpicture}[remember picture, overlay]
      \draw[-{Stealth}, thick, red]
        ([xshift=2pt, yshift=4pt]{pic cs:startF})
        to[out=0, in=180, looseness=1.5]
        ([xshift=2pt, yshift=4pt]{pic cs:endL});
    \end{tikzpicture}

\begin{tabular}{r|rrrrr}
$i$ & $\ICA{1}$ & $\CA{1}$ & $F$ & $L$ & $\LCPinf{1}$ \\
\hline
1  & 1  & 11 & \$ & 0 & 0 \\
2  & 2  & 10 & 0  & 0 & 0 \\
3  & 7  & 6  & 3  & 0 & 1 \\
4  & 5  & 1  & 4  & 1 & 1 \\
5  & 9  & 8  & 1  & 0 & 1 \\
6  & 3  & 3  & 1  & 0 & 2 \\
7  & 8  & 9  & 0  & 1 & 1 \\
8  & 11 & 5  & 0  & 0 & 2 \\
9  & 6  & 7  & 0  & 3 & 2 \\
10 & 10 & 2  & 0  & 4 & 3 \\
11 & 4  & 4  & 0  & 1 & 2 \\
\end{tabular}
    \end{minipage}
\begin{minipage}{0.48\linewidth}
\begin{tabular}{r|rrrrr}
$i$ & $\ICA{1}$ & $\CA{1}$ & $F$ & $L$ & $\LCPinf{1}$ \\
\hline
1  & 1  & 16 & \$ & 0  & 0 \\
2  & 2  & 15 & 0  & 0  & 0 \\
3  & 9  & 11 & 3  & 0  & 1 \\
4  & 6  & 6  & 4  & 0  & 1 \\
5  & 12 & 1  & 4  & \$ & 1 \\
6  & 3  & 13 & 1  & 0  & 1 \\
7  & 10 & 8  & 1  & 0  & 2 \\
8  & 15 & 3  & 1  & 0  & 3 \\
9  & 7  & 14 & 0  & 1  & 1 \\
10 & 13 & 10 & 0  & 0  & 2 \\
11 & 4  & 5  & 0  & 0  & 2 \\
12 & 11 & 12 & 0  & 3  & 2 \\
13 & 16 & 7  & 0  & 4  & 3 \\
14 & 8  & 2  & 0  & 4  & 4 \\
15 & 14 & 9  & 0  & 1  & 2 \\
16 & 5  & 4  & 0  & 1  & 3 \\
\end{tabular}
\end{minipage}

\caption{Insertion of the sequence $[1,5,3,4,2]$ with the \twoFold{} (left) or threefold (right) construction into an empty ceBWT\@. 
        The rows of the second part that will stay in the BWT are highlighted in blue.
        The red arrow indicates a mismatch of the $F$ and $L$ values of the first and the last entry of the sequence, which need to be made equal.
        The algorithm does so by identifying the mismatching value $3$ in $F$ with the value~$4$ in $L$ and adjusts $3$ to $4$ in $F$.
        Right: The threefold construction avoids this mismatch by providing sufficient context.
    }
    \label{fig:folding-example}
\end{figure}

We implement multiple variants for constructing the dynamic index. One variant follows the fourfold construction of the original paper; the other variants reduce the replicated input used during construction and add multithreading.
The input sequence replication is required because comparing two conjugates of length $n$ in the $\omega$-preorder may require $3n$ symbols, as shown in~\cite[Lemma 3.3]{osterkamp26cebwt}. 
Replicating a sequence fourfold therefore gives every conjugate a sufficient comparison range in the resulting linear sequence. The replicated sequence is used only during construction; the retained index structures $L$, $F$, and $\mathrm{LCP}^{\infty}$ contain only the entries corresponding to the actual conjugates of the input text.

The reduced constructor \twoFold{} uses two copies for the initial construction and three copies when computing comparison values for merging into an existing index. 
This reduction is an engineering optimization rather than a new theoretical construction. 
It is guarded by a consistency check on the multiplicities represented by $L$ and $F$ in the single index construction.
Such a consistency check is required, as can be seen by the single index of the example sequence $S = [1, 5, 3, 4, 2]$ presented in~\cref{fig:folding-example}. 
The discrepancy affects only the F-value $\textup{F}[i]$ with $i$ such that $\textup{CA}[i] = \left|S\right|+1 = 6$, and stems from the auxiliary symbol $\$$ and the length we compare to. 
We obtain the correct F-value from $\textup{F}[i]$ with $i$ such that $\textup{CA}[i] = 1$, and overall correctness of the optimized single index construction (up to tie-breaks in the conjugate array) is an immediate consequence of~\cref{lemma:rotaencodingstrongmatch} presented below since conjugates of a given single string have equal length.

\begin{lemma}[{\cite[Lemma~A.1]{osterkamp26cebwt}}]\label{lemma:rotaencodingstrongmatch}
	Let $V, U \in \Sigma^+$. 
	Then $\pde{V} = \pde{U}$ if and only if $V \approx U$.
\end{lemma}

For the comparison with length $3n$ instead of $4n$ during insertion, the correctness of the implementation follows under the requirement that the text $R$ subject to insertion is given by one of its rotations starting with an occurrence of the smallest symbol it contains. 
We can achieve that for general texts $R$ if we rotate it such that its smallest symbol appears at the first position $\textup{Rot}(R,i)[1]$ and then build the index for $\rta{R}{i}^3\cdot\$$.
The technical details are as follows, where we borrow two more results from literature to establish the correctness in \cref{lem:equality2z+}. 

\begin{lemma}[Weak Periodicity Lemma]\label{lem:weakperiodicity}
	Let $p$ and $q$ be two periods of a string $X$. If $p+q \leq \left|X\right|$, then $\gcd(p,q)$ is also a period of $X$.
\end{lemma}

\begin{lemma}[{\cite[Section~2.4]{park20patternsperiods}}]\label{lem:encctmatch}
		Let $U,V \in (\Sigma \cup \{\$\})^*$. 
		The following statements are equivalent.
		\begin{enumerate}[(i)]
			\item $U \approx V$.
			\item $\pde{U} = \pde{V}$.
			\item $\left|U\right| = \left|V\right|$ and $U[i..j] \approx V[i..j]$ for each $1 \leq i \leq j \leq \left|U\right|$.
			\item $\left|U\right| = \left|V\right|$ and $\pde{U}[i..j] = \pde{V}[i..j]$ for each $1 \leq i \leq j \leq \left|U\right|$.
		\end{enumerate}
	\end{lemma}

    \begin{lemma}\label{lem:equality2z+}
		Let $V, U \in (\Sigma \cup \{\$\})^+$ and $z = \max\{\left|V\right|, \left|U\right|\}$. 
		Then $V \cteq U$ if and only if $\pde{V^{\omega}[..2z+y-1]}  = \pde{U^{\omega}[..2z+y-1]}$, where $y = \min\{\max\{ 1 \leq i \leq \left|U\right|\mid \pde{U}[i] = \infty\},\max\{ 1 \leq j \leq \left|V\right|\mid \pde{V}[j] = \infty\}\}$.
	\end{lemma}

    \paragraph{Proof}
    
		Without loss of generality, $\left|U\right| = z$.
		Let $\left|V\right|= i$, $\left|\wurz{\pde{V}}\right| = j$ and $\left|\wurz{\pde{U}}\right| = k$.
		\begin{description}
			\item[$(\Leftarrow)$] Assume $\pde{V^{\omega}[..2z+y-1]}  = \pde{U^{\omega}[..2z+y-1]}$.
			If $i = z$, then the statement follows from Lemma~\ref{lemma:rotaencodingstrongmatch} and $ \pde{V}   =\pde{V^{\omega}[..z]} = \pde{U^{\omega}[..z]} = \pde{U}$.
			Thus, assume $i < z$.
			Then, on the one hand,
			\[
			\begin{split}
				\rpde{\rta{V}{y}}^\omega[..2z-1] &= \pde{V^{\omega}[..2z+y-1]}[y+1..] \\
				&= \pde{U^{\omega}[..2z+y-1]}[y+1..] = \rpde{\rta{U}{y}} \cdot \rpde{\rta{U}{y}}[..z-1]
			\end{split}
			\]
			by Lemma~\ref{lem:encctmatch}.
			Since the rotational parent distance encoding is commutative with left rotations, $j$ is a period of $\rpde{\rta{V}{y}}$ and $z$ is a period of $\rpde{\rta{U}{y}}$.
			Consequently, both $z$ and $j$ are periods of $\rpde{\rta{U}{y}} \cdot \rpde{\rta{U}{y}}[..z-1]$.
			Since $j + z \leq 2z-1 = \left|\rpde{\rta{U}{y}} \cdot \rpde{\rta{U}{y}}[..z-1]\right|$, we can apply Lemma~\ref{lem:weakperiodicity} and find that $\gcd(j,z)$  is a period of $\rpde{\rta{U}{y}} \cdot \rpde{\rta{U}{y}}[..z-1]$.
			As $\gcd(j,z)$ divides $z = \left|\rpde{\rta{U}{y}}\right|$, $\rpde{\rta{U}{y}}$ can be formed by repeating $\rpde{\rta{U}{y}}[..\gcd(j,z)]$ an integral number of times, which implies $\gcd(j,z) \geq k$, i.e., $i \geq j \geq k$. 
			On the other hand,
			\[
			\begin{split}
				\rpde{\rta{V}{y}} \cdot \rpde{\rta{V}{y}} &= \pde{V^{\omega}[..2z+y-1]}[y+1..y+1+2i] \\ 
				&= \pde{U^{\omega}[..2z+y-1]}[y+1..y+1+2i] = \rpde{\rta{U}{y}}^\omega[..2i]
			\end{split}
			\]
			by Lemma~\ref{lem:encctmatch}. 
			Since the rotational parent distance encoding is commutative with left rotations, $k$ is a period of $\rpde{\rta{U}{y}}$ and $j$ is a period of $\rpde{\rta{V}{y}}$.
			As $k \leq i$, $k$ is a period of $\rpde{\rta{V}{y}} \cdot \rpde{\rta{V}{y}}$ in addition to $i$.
			Then $k + i \leq 2i = \left|\rpde{\rta{V}{y}} \cdot \rpde{\rta{V}{y}}\right|$ and Lemma~\ref{lem:weakperiodicity} imply that $\gcd(k,i)$ is a period of $\rpde{\rta{V}{y}} \cdot \rpde{\rta{V}{y}}$.
			As $\gcd(k,i)$ divides $i = \left|\rpde{\rta{V}{y}}\right|$, $\rpde{\rta{V}{y}}[..\gcd(k,i)]$ can be repeated an integral number of times to form $\rpde{\rta{V}{y}}$, which implies $\gcd(k,i) \geq j$.
			
			Consequently, $k \geq j$, and therefore $\left|\wurz{\rpde{V}}\right| = j = k = \left|\wurz{\rpde{U}}\right|$.
			In particular, $\wurz{\rpde{V}} = \pde{V^{\omega}[..2z]}[j+1..2j]  = \pde{U^{\omega}[..2z]}[k+1..2k] = \wurz{\rpde{U}}$, i.e., $V \cteq U$.
			\item[$(\Rightarrow)$] Let $V \cteq U$.
			Assume $\pde{V^{\omega}[..2z+y-1]} \neq \pde{U^{\omega}[..2z+y-1]}$ with $x$ minimal such that $\pde{V^{\omega}[..2z+y-1]}[x] \neq \pde{U^{\omega}[..2z+y-1]}[x]$. 
			If $\max \{\pde{V^{\omega}[..2z+y-1]}[x], \pde{U^{\omega}[..2z+y-1]}[x]\} < \infty$, then $\rta{\wurz{\rpde{V}}}{x-1}[1] = \pde{V^{\omega}[..2z+y-1]}[x] \neq \pde{U^{\omega}[..2z+y-1]}[x] = \rta{\wurz{\rpde{U}}}{x-1}[1]$, which contradicts $V \cteq U$.
			Hence, and without loss of generality, assume $\pde{V^{\omega}[..2z+y-1]}[x] = \infty$.
			Then $\pde{U^{\omega}[..2z+y-1]}[x] < \infty$, $x \leq \left|\wurz{\rpde{V}}\right|$, and consequently $\wurz{\rpde{U}}[x] = \pde{U^{\omega}[..2z+y-1]}[x] < x \leq \wurz{\rpde{V}}[x]$, a contradiction to $V \cteq U$.
			Thus, $\pde{V^{\omega}[..2z+y-1]}  = \pde{U^{\omega}[..2z+y-1]}$. 
            \hfill $\blacksquare$
		\end{description}

Thus, with the adjustment to the insertion in the implementation by rotating the new text to a conjugate starting with an occurrence of its smallest appearing symbol,
we have a theoretical foundation for the implementation of the threefold variant and \twoFold{}.

\subsection{Parallelism}\label{sec:parallelism}

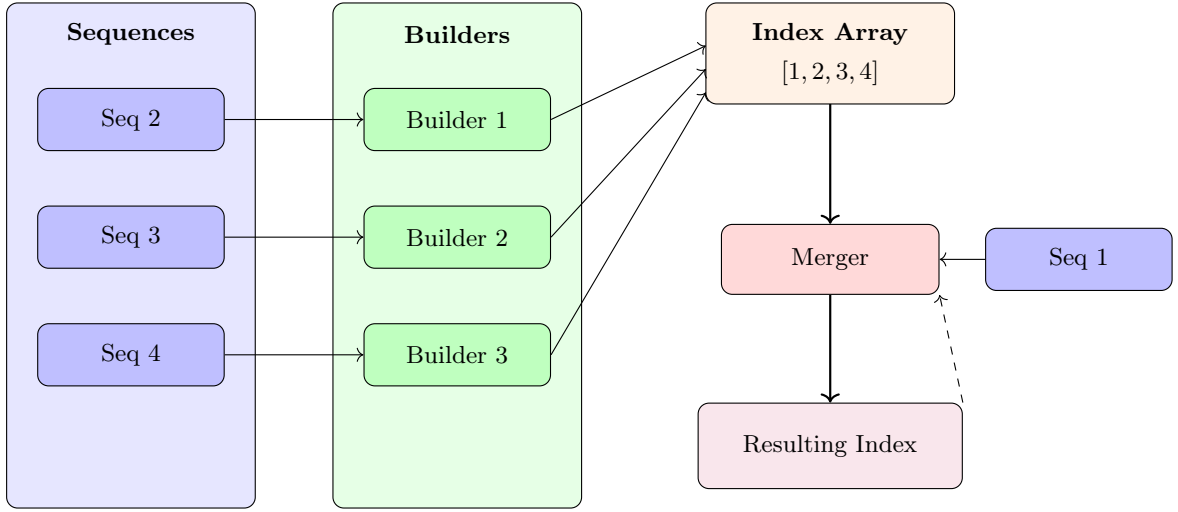
\begin{figure}[t]
    \centering
    \resizebox{\textwidth}{!}{\centering
    \begin{tikzpicture}[font=\small, scale=1.0, transform shape]
\draw[rounded corners, fill=blue!10] (0,0) rectangle (3.2,6.5);
    \node[font=\bfseries\footnotesize] at (1.6,6.1) {Sequences};
    \node[draw, rounded corners, fill=blue!25, minimum width=2.4cm, minimum height=0.8cm] (s2) at (1.6,5.0) {\footnotesize Seq 2};
    \node[draw, rounded corners, fill=blue!25, minimum width=2.4cm, minimum height=0.8cm, below=0.7cm of s2] (s3) {\footnotesize Seq 3};
    \node[draw, rounded corners, fill=blue!25, minimum width=2.4cm, minimum height=0.8cm, below=0.7cm of s3] (s4) {\footnotesize Seq 4};
\draw[rounded corners, fill=green!10] (4.2,0) rectangle (7.4,6.5);
    \node[font=\bfseries\footnotesize] at (5.8,6.1) {Builders};
    \node[draw, rounded corners, fill=green!25, minimum width=2.4cm, minimum height=0.8cm] (b1) at (5.8,5.0) {\footnotesize Builder 1};
    \node[draw, rounded corners, fill=green!25, minimum width=2.4cm, minimum height=0.8cm, below=0.7cm of b1] (b2) {\footnotesize Builder 2};
    \node[draw, rounded corners, fill=green!25, minimum width=2.4cm, minimum height=0.8cm, below=0.7cm of b2] (b3) {\footnotesize Builder 3};
\draw[rounded corners, fill=orange!10] (9.0,5.2) rectangle (12.2,6.5);
    \node[font=\bfseries\footnotesize] at (10.6,6.1) {Index Array};
    \node[font=\footnotesize] at (10.6,5.6) {$[1,2,3,4]$};
\node[draw, rounded corners, fill=red!15, minimum width=2.8cm, minimum height=0.9cm]
        (m) at (10.6,3.2) {\footnotesize Merger};
\node[draw, rounded corners, fill=blue!25, minimum width=2.4cm, minimum height=0.8cm]
        (s1) at (13.8,3.2) {\footnotesize Seq 1};
\node[draw, rounded corners, fill=purple!10, minimum width=3.4cm, minimum height=1.1cm]
        (r) at (10.6,0.8) {\footnotesize Resulting Index};
\draw[->] (s2.east) -- (b1.west);
    \draw[->] (s3.east) -- (b2.west);
    \draw[->] (s4.east) -- (b3.west);
    \draw[->] (b1.east) -- (9.0,5.95);
    \draw[->] (b2.east) -- (9.0,5.65);
    \draw[->] (b3.east) -- (9.0,5.35);
    \draw[->, thick] (10.6,5.2) -- (m.north);
    \draw[->] (s1.west) -- (m.east);
    \draw[->, thick] (m.south) -- (r.north);
    \draw[->, dashed] (r.north east) -- (m.south east);
    \end{tikzpicture}
    }
    \caption{Threaded construction model of \twoFold{}. Worker builders construct partial indices for later sequences, while the main builder merges them with the existing index in sequence order.}
    \label{fig:threaded-construction-model}
\end{figure}

In the multithreading approach, a configurable number of threads construct the index as follows: one thread acts as the main constructor and merger, building the index for the first sequence and subsequently receiving partial indices from the remaining threads to merge. This design effectively implements a sequential task queue, ensuring that merges are applied in the correct order while computations are performed in parallel; see \cref{fig:threaded-construction-model}. The first sequence is handled by the merger thread itself, so that the global dynamic index is initialized before worker-produced partial indices are integrated. For very small input collections, where parallel construction has no practical benefit and worker-queue edge cases dominate, the implementation falls back to the single-threaded construction path. This variant also uses \twoFold{} described above. These optimizations aim to reduce the overall runtime. General observations regarding the effectiveness and impact of these optimizations are discussed further in \cref{chapter:benchmarks}.

The dynamic implementation uses a deterministic tie-breaking rule for
CTM-equivalent sequences.
The original theoretical construction~\cite{osterkamp26cebwt} inserts a new sequence below an already present CTM-equivalent sequence.
Our implementation uses the opposite convention: when a new sequence is
CTM-equivalent to sequences already present in the index, it is inserted above them.
This choice only fixes the internal order among CTM-equivalent entries.
Since all these entries have the same Cartesian tree, the convention does not
affect the represented CTM equivalence classes or the answers returned by
\texttt{count}; it only has to be applied consistently during construction and updates.

The threaded constructor required additional engineering safeguards. In particular, the queue used for partial indices must not be persistent across independent benchmark repetitions, and the merge thread must not wait for worker output before an initial index has been constructed. We therefore keep the synchronization state local to each construction call, merge partial indices according to their sequence numbers, and use a single-threaded fallback for collections that are too small to benefit from worker threads. These choices do not change the index semantics, but they are necessary to make the parallel constructor deterministic and reproducible under repeated benchmark execution.

\subsection{Validation}
\label{sec:validation}
We validate the correctness of the implementation through multiple tests on smaller input sequences. For this purpose, we use the GoogleTest library~\cite{google_testing_framework_googletest_2026}. In addition, all critical helper functions are tested for correctness. We provide a script in the repository to compute the test coverage.

We test all construction variants against the naive implementation on a range of inputs, including edge cases such as sequences with duplicate values and patterns longer than the text. The tests compare the computed count intervals and match counts with the naive PDE-based search, and also check circular wraparound cases. These tests validate that the implementation handles the boundary conditions relevant for the benchmarked count queries.

\subsection{Implementation Details}
\label{chapter:implementationdetails}

We use the libraries SDSL v2.1.1~\cite{gog14sdsl} and Dynamic v1.0-alpha.1~\cite{prezza17dynamic}. 

From SDSL, we primarily use static bit vectors with rank and select support. Specifically, \texttt{bit\_vector} stores a bit sequence $B \in \{0,1\}^n$ in $64\lceil n/64 + 1\rceil$ bits of memory, with \texttt{rank\_support\_v} requiring $0.25n$ bits of additional memory per instance for $\mathcal{O}(1)$ rank queries and \texttt{select\_support\_mcl} requiring at most $0.2n$ bits of additional memory per instance for $\mathcal{O}(1)$ select queries~\cite{gog_sdsl_nodate}. 

From Dynamic, we use the dynamic succinct bit vector \texttt{suc\_bv}, which stores $B \in \{0,1\}^n$ in $n + o(n)$ bits of memory and supports rank, select, access and insert operations in $\mathcal{O}(\log n)$ time, as well as dynamic wavelet trees. Let $S \in \Sigma^n$ be a sequence with zero-order entropy $H_0$. The dynamic wavelet trees store $S$ in $n(H_0 + 1)(1 + o(1)) + \mathcal{O}(|\Sigma| \log n)$ bits of memory and support rank, select, insert and access over sequences on average in $\mathcal{O}((H_0 + 1)\log n)$ time~\cite{prezza17dynamic}.
Dynamic construction has a significant impact on runtime~\cite{navarro14dynamic}. Therefore, we compared Dynamic to another library~\cite{miti7_waveletmatrix_2026} for our implementation, but its runtime was not significantly better and the memory consumption was considerably higher.

Since no publicly available implementation of the ceBWT exists, direct comparisons with a reference implementation using the same data structure are not possible. 
Instead, we validate the ceBWT count results against a naive implementation. The naive implementation computes matches by explicitly constructing and comparing PDEs for every candidate occurrence, including circular occurrences where applicable. This implementation is not optimized for efficiency and is used solely to check that the properties and behavior described in the original paper are implemented correctly.

\section{Benchmark Methodology}
\label{chapter:methodic}
\subsection{Hardware Environment}
The implementation is written in C++20 and was compiled with g++~14.2.0 using the flags \texttt{-Ofast -march=native -mtune=native -flto -DNDEBUG}
on Debian Trixie.
The benchmark machine used an AMD Ryzen 9 5900XT CPU and
\texttt{128}~GiB of main memory.
CPU frequency scaling was fixed to the performance governor during all benchmark runs.
Unless stated otherwise, benchmarks were run on this machine with the same compiler, compiler flags and input files.

\subsection{Artifact Availability and Reproducibility}
\label{section:artifact-availability}

The implementation, benchmark drivers and plotting scripts are available at
\url{https://github.com/koeppl/cebwt}.
All results reported in this article were generated from commit
\texttt{0071b0b} using tag \texttt{TCS26}.

The benchmark inputs, generated pattern sets, precomputed dynamic and static
indices, raw measurement files, and plotting data are available in the data
repository at \url{https://github.com/koeppl/cebwt_data}.
The raw benchmark output is stored under \texttt{benchmark/results}. The files
contain the individual repetitions used to compute the medians and
interquartile ranges shown in the plots.
The scripts used to generate the datasets and figures are included in the
repository, and the corresponding commands are documented in the repository
README.

The generated MIDI subsets, genomic subsequences, and pattern files are included
in the data repository, so the reported experiments can be rerun without resampling the inputs.
The NCBI accessions used for the genomic inputs are listed in
\cref{tab:genomes}; the exact download commands and preprocessing scripts
are included with the artifact.

For convenience, the artifact also provides a Docker-based reproduction path.
The repository contains a benchmark Dockerfile and a
helper script, \texttt{docker/build\_and\_run\_benchmarks.sh}. 
The script
initializes the \texttt{cebwt\_data} submodule, builds the Docker image, runs the
benchmark suite inside the container, and bind-mounts the result directory so
that the generated CSV files and plots are written back to the host under
\texttt{benchmark/results}.  
The Docker setup fixes the operating-system environment and the software
packages used to compile the benchmark executables.  It does not remove
hardware-dependent variation, so the absolute running times reported below are
expected to depend on the processor and memory system; however, the same scripts
recreate the raw files and figures from the supplied inputs.

\subsection{Datasets}
\label{section:dataset}
All datasets used in this work are provided in our data repository. To cover diverse input types, we evaluate the benchmarks on MIDI files and genomic sequences across different application scenarios.

We randomly extracted 203 MIDI files from the collection introduced in~\cite{tian_xmusic_2025} and tokenized them using the \texttt{pretty\_midi} library~\cite{raffel_intuitive_2014}. Each file is represented as a sequence of integer values in the range $[0..127]$. The tokenizer is available in the dataset repository. The sequences range from 109 to 7{,}079 tokens, with a median length of 1{,}470 tokens and a total of 363{,}164 tokens.

For the genomic datasets, we define three distinct groups:
\begin{enumerate}
    \item 12 subsequences of the \textit{Escherichia coli} K-12 MG1655 genome, each 24{,}000 bases long,
    \item 16 real genomes ranging from a few thousand to 50{,}000 bases, where segmented genomes such as Mammalian orthoreovirus contribute multiple sequences and
    \item 16 subsequences derived from the highly repetitive~\cite{sasaki_map-based_2005} rice genome (\textit{Oryza sativa}).
\end{enumerate}
All real genomes were downloaded from the NCBI database~\cite{national_center_for_biotechnology_information_ncbi_2025}; their accession numbers are listed in \cref{tab:genomes}. The \textit{E.~coli} genome was chosen to provide a representative real genome structure. Since the rice and \textit{E.~coli} genomes are large and not directly used in their entirety, they are not included in the repository. All download queries are documented in the README for reproducibility.

Each subsequence in group~(3) matches the length of one sequence from group~(2) and carries the same name with an additional \texttt{\_sim} suffix.
To ensure that only highly repetitive subsequences are retained, we generated 3{,}000 candidate subsequences per target length. The candidate with the lowest $k$-mer entropy~\cite{FORET2009539}, 
computed over its PDE, was selected, as this structure is central to CTM as explained in \cref{section:CTS}. This pairing allows us to directly compare performance on real and repetitive sequences of identical length, isolating the effect of repetitiveness on our index. For extra safety, we also computed the runs of all sequences with~\cite{koeppl_stringology-rust_nodate}. The Python scripts used and all measured repetitiveness scores are available in the repository. 

Patterns were generated using \texttt{PatternGenerator} and are provided in the repository under the \texttt{extras} folder. For the MIDI dataset, we generated 50 patterns for each length from 25 to 1{,}000 in steps of 25. For the genomic dataset, we generated 50 patterns of length 1{,}000. To evaluate scalability on genomic data, shorter lengths are obtained during benchmarking by truncating each pattern to the desired length, ensuring structural consistency across different lengths. 

Precomputed dynamic and static index structures are included in the repository to allow benchmarking without rebuilding the index in every scenario. These were generated using \texttt{DataGenerator}. 
\cref{tab:benchmark-dataset-overview} summarizes the dataset and pattern groups used in the experiments.

\begin{table}[t]
    \centering
    \small
    \textbf{Datasets \& Patterns}:
    \begin{center}
    \begin{tabular}{|p{2.3cm}|p{7.2cm}|p{4.5cm}|}
            \hline
            & Genomes & Music \\
            \hline
        Datasets & 12 subsequences of \textit{E.~coli} \newline 16 rice-derived repetitive subsequences \newline 16 full genomes & 203 MIDI files
            \\
            \hline
            Patterns & $50 \times 1000$ bases & $50 \times 25$--$1000$ tokens\\
            \hline
        \end{tabular}
    \end{center}
    \caption{Dataset and pattern overview used for the benchmark suite. The genomic experiments use fixed-length genomic patterns, whereas the MIDI experiments cover a range of token-pattern lengths.}
    \label{tab:benchmark-dataset-overview}
\end{table}

\begin{table}[htbp] 
  \centering        
  \begin{tabular}{|l|l|}
    \hline
    \textbf{Name} & \textbf{Accession} \\
    \hline
    \textit{Escherichia coli} K-12 MG1655 & GCF\_000005845.2 \\
    Tunavirus T1 & GCA\_008227805.1 \\
    Salavirus & GCF\_002755075.1 \\
    SARS-CoV-2 & GCF\_009858895.2 \\
    Mammalian orthoreovirus & GCF\_006298385.1 \\
    Adenovirus & GCA\_006402115.1 \\
    MERS-CoV & GCF\_000901155.1 \\
    Lambda 1H12 & GCF\_902006465.1 \\
    Rice (\textit{Oryza sativa}) & GCF\_034140825.1 \\
    \hline
  \end{tabular}
  \caption{Genome datasets downloaded from NCBI Datasets \cite{national_center_for_biotechnology_information_ncbi_2025}.}
  \label{tab:genomes}
\end{table}

\subsection{Experimental Design}
We benchmark the performance of our implementation by measuring construction throughput (symbols/second), search runtime, search throughput (queries/second), peak memory consumption (bytes/symbol), and index size (bytes/symbol).
\cref{tab:benchmark-measurement-matrix} records which quantities are measured for each benchmarked phase.

\begin{table}[t]
    \centering
    \small
    \textbf{What is measured?}
    \begin{center}
        \begin{tabular}{|p{4.3cm}|p{3.5cm}|p{3.5cm}|}
            \hline
            & Runtime & Memory \\
            \hline
            Construction & \checkmark & \checkmark \\
            \hline
            Conversion & \checkmark & \checkmark \\
            \hline
            Index size & & \checkmark \\
            \hline
            Pattern search & \checkmark & \\
            \hline
            Number of matches & & \\
            \hline
        \end{tabular}
    \end{center}
    \caption{Measurement matrix for the benchmark suite. Construction and conversion are evaluated in both runtime and memory, index size is evaluated as memory, and pattern search is evaluated as runtime together with the number of reported matches.}
    \label{tab:benchmark-measurement-matrix}
\end{table}

All benchmarks for our ceBWT use the \texttt{CAStorage}, as it gives the best balance between supported operations and memory usage.

For index construction, we measure throughput (symbols/second) and peak memory usage (bytes/symbol). For pattern matching, 
we distinguish two evaluation modes: when comparing algorithms across patterns of varying lengths, we report total runtime; when evaluating scalability over a single pattern-length axis with a fixed set of queries per length, we report throughput (queries/second). In both cases, we additionally report the number of found matches. For the KMP-based baselines, we also report memory usage, since they allocate memory directly prior to the search itself. We additionally report the memory footprint of auxiliary structures, namely the conjugate array and endpoint vector for the selected storage, the bit vector $e$ for the dynamic index and the $\mathrm{LCP}^{\infty}$ wavelet tree for the static index, as described in \cref{section:implementationdifferences}.

To obtain stable runtime measurements, we perform 100 warm-up runs and 300 repetitions for most benchmarks. Exceptions are the dynamic index construction, where we use 5 warm-up runs and 20 repetitions due to its longer execution time, and the KMP baselines, where we use 20 warm-up runs and 20 repetitions. We report the median runtime together with the interquartile range. The latter is available in our repository and was used to inspect outliers. For memory measurements, we use 1 warm-up run and 1 repetition, as we observed no variation across runs after the first. The dynamic index construction is an exception, where we use 1 warm-up run and 20 repetitions due to its threading logic.

As baselines for exact string matching, we use SDSL~\cite{gog14sdsl}, r-index~\cite{gagie18bwt,gagie20fully}, and RLCSA~\cite{novak_rlcsa_2010}. We additionally include the compact order-preserving FM-index of Decaroli et al.~\cite{decaroli19compact}; this baseline is denoted by OPM in the experimental plots. 
As baselines for CTM, we use two KMP-based variants following Park et al.~\cite{park19cartesiantreematching}: one based on PDE and one based on CTS with the auxiliary array $D$ described in \cref{section:CTS}.
Like the ceBWT, our KMP implementations support cyclic pattern matching.
Conceptually, we can achieve that for any online pattern matching algorithm by appending the first $m-1$ symbols of the text to its end, where $m$ is the pattern length.
In practice, we achieve the same effect by using modular arithmetic to access the text symbols, so no actual concatenation is needed.
All baselines are evaluated on the same input sequences and patterns. To avoid I/O overhead, the KMP baselines were evaluated using the in-memory vectors API rather than their default file-based input.
\cref{tab:benchmark-baselines} summarizes the baseline algorithms by matching model.

\begin{table}[t]
    \centering
    \small
    \textbf{Benchmarked baseline families}:
    
    \begin{center}
    \begin{tabular}{|p{2.8cm}|p{4.3cm}|p{3.0cm}|p{3.0cm}|}
            \hline
            Matching Type & Exact Matching & OPM & CTM \\
            \hline
            Algorithm & 
            SDSL FM-index~\cite{gog14sdsl} \newline 
            RLCSA~\cite{novak_rlcsa_2010} \newline 
            r-index~\cite{gagie20fully}\newline ~ 
            & 
            OPM-FM~\cite{decaroli19compact} \newline
            &
            KMP-PDE \newline
            KMP-CTS \newline \\
            \hline
        \end{tabular}
    \end{center}
    \caption{Benchmarked baseline families grouped by matching model. Exact string matching is represented by SDSL, RLCSA, and the r-index; order-preserving matching is represented by the compact OPM-FM index; Cartesian-tree matching is represented by KMP-based baselines using PDE and CTS.}
    \label{tab:benchmark-baselines}
\end{table}

For the threaded dynamic constructor, we evaluate one to four worker threads in \cref{fig:ThreadedBM}; the setting is further described in \cref{section:genomedata}. 
When a single threaded configuration is used in a benchmark, the number of workers is stated together with the corresponding experiment. The thread-scaling benchmark and additional phase profiling show that the current implementation improves over the sequential constructor, but is limited by the ordered merge phase and does not obtain a consistent throughput gain beyond two workers.

We measure runtime using \texttt{std::chrono} from the C++ standard library. Memory consumption is tracked using \texttt{malloc\_count}~\cite{bingmann_malloc_count_2014}, which records the maximum heap allocation during execution excluding memory allocated prior to the measurement. Since the threaded constructor creates worker threads during the measured run, the allocator instrumentation must also be safe during \texttt{pthread\_create}. We therefore use a thread-safe version of \texttt{malloc\_count}: the global allocation counters are updated atomically, peak memory is maintained by an atomic compare-and-swap loop, and \texttt{calloc} is implemented without recursively calling the instrumented \texttt{malloc} wrapper. This avoids deadlocks in the allocator interposition layer and allows the same heap-memory metric to be used for both sequential and threaded construction. The memory footprint of index structures is obtained via the \texttt{bit\_size} or \texttt{serialize} methods provided by the respective libraries. All results are available in our repository under \texttt{benchmark/results}. In the generated CSV files, we report the runtime in seconds and all memory values in kibibytes. For the analysis, memory is converted to bytes per symbol to allow reasonable ranges for comparison.

To integrate the string pattern matching baselines into our benchmarking framework, we removed print statements from their implementations and created wrapper classes. For the exact string-matching baselines, wrapper correctness was checked by comparing the returned match counts across independent exact indexes on the same inputs.

For genomic pattern matching, we search 50 patterns for each length, where shorter lengths are obtained by the truncation mechanism described in \cref{section:dataset}. For the MIDI dataset, all available 2{,}000 patterns are searched at once, covering 50 patterns for each of the 40 lengths.

\subsection{Benchmark Structure}
We organize our benchmarks into two major groups: \textit{internal} benchmarks on genomic data using only our implementation and \textit{external} benchmarks on MIDI data comparing our implementation against baseline algorithms. We present results as bar charts, line plots, and tables.

The internal group is divided into four parts corresponding to the dataset groups defined in \cref{section:dataset}. In the first part, we evaluate index construction on \textit{E.~coli} subsequences across different sequence counts and lengths, covering all dynamic construction variants, static construction, and index sizes including additional structures. The second part evaluates the full real genomes, reporting construction time across varying thread counts and input orderings, and pattern matching scaling. The third part directly compares real and repetitive genomes of identical length, reporting construction time, index sizes, and pattern matching results. The fourth part compares measured values with fitted asymptotic trends using the least-squares method~\cite[Section 15.4]{press_numerical_2007}.

The external group first compares construction time, index sizes, and pattern matching against exact string-matching baselines and the OPM-FM index. We then compare our implementation against the KMP-based CTM baselines separately, as they perform pattern search without constructing a persistent index and therefore require a different evaluation setup. We conclude with a discussion of the number of found matches over all algorithms.\section{Analysis and Discussion}
\label{chapter:benchmarks}
\subsection{Internal Benchmarks on E.~coli Subsequence Data}
\label{section:EColiBM}
\begin{figure}[htbp]
    \centering
\begin{subfigure}[b]{0.45\textwidth}
        \centering
        \includegraphics[width=\textwidth]{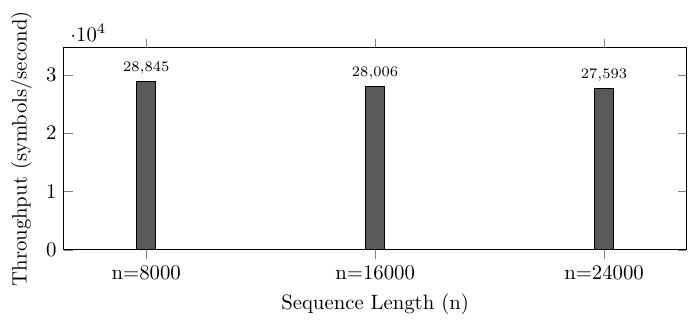}
        \caption{Throughput (symbols/second)}
        \label{fig:builderrun}
    \end{subfigure}
    \hfill
    \begin{subfigure}[b]{0.45\textwidth}
        \centering
        \includegraphics[width=\textwidth]{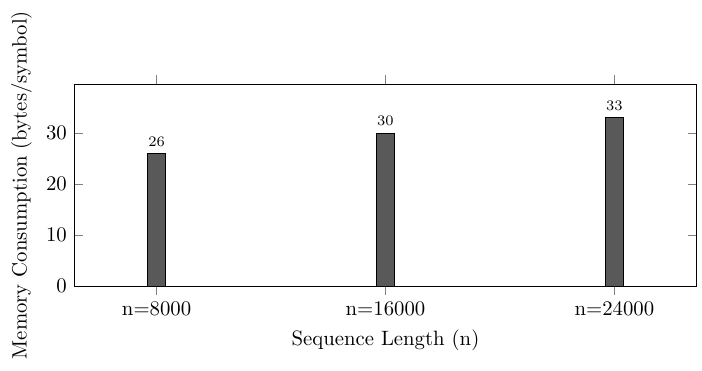}
        \caption{Memory consumption (bytes/symbol)}
        \label{fig:buildermem}
    \end{subfigure}
    \caption{Comparison of throughput (left) and memory consumption per symbol (right) during dynamic index construction at constant total symbol count of 96{,}000 across varying sequence lengths on our \textit{E.~coli} subsequences.}
    \label{fig:builder_overview}
\end{figure}

\cref{fig:builder_overview} shows the impact of varying the sequence length and sequence count on throughput and memory consumption during dynamic index construction. We tested three configurations: 12 sequences of length 8{,}000, 6 sequences of length 16{,}000, and 4 sequences of length 24{,}000, each totaling 96{,}000 symbols. The plot in \cref{fig:builderrun} shows the throughput in symbols per second, while the plot in \cref{fig:buildermem} shows the memory consumption in bytes per symbol.

Throughput decreases with increasing sequence length, while memory consumption per symbol grows.
Specifically, the 8{,}000 symbol configuration achieves a throughput of 29{,}210 symbols per second, compared to 28{,}238 for length 16{,}000 and 27{,}967 for length 24{,}000. The absolute difference decreases with increasing length, dropping by 972 between the first two configurations and by 271 between the last two.
For memory consumption, the values are 26, 30, and 33 bytes per symbol. The difference decreases from 4 bytes per symbol between the first two configurations to 3 bytes per symbol between the last two.

This behavior suggests that sequence length has a larger impact on construction cost than the number of sequences.
Since \twoFold{} requires the twofold input length extension for the initial index computation of a single text and the additional tripling for extending the existing index, the introduced overhead grows with sequence length and exceeds the cost of repeating the process for more, shorter sequences.
Threading reduces the impact of the former, but computation of comparison values remains the dominant bottleneck. Memory behavior reflects a similar pattern, as the computation of comparison values requires more memory per symbol for longer sequences. Overall, the results indicate that throughput decreases with increasing sequence length, with diminishing marginal loss; at equal total symbol count, shorter sequences achieve higher throughput.

\begin{figure}[htbp]
    \centering
    \begin{subfigure}[b]{0.45\textwidth}
        \centering
        \includegraphics[width=\textwidth]{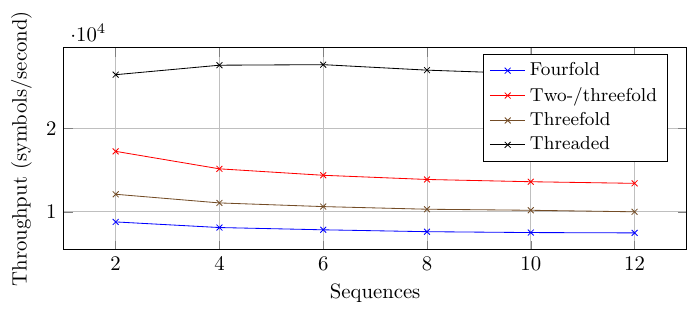}
        \caption{Throughput (symbols/second)}
        \label{fig:builderConstructorsRun}
    \end{subfigure}
    \hfill
    \begin{subfigure}[b]{0.45\textwidth}
        \centering
        \includegraphics[width=\textwidth]{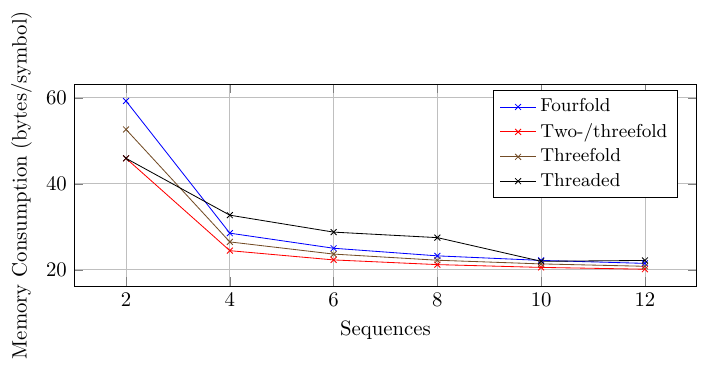}
        \caption{Memory consumption (bytes/symbol)}
        \label{fig:builderConstructorsMem}
    \end{subfigure}
    \caption{Comparison of throughput (left) and memory consumption per symbol (right) during dynamic index construction across four constructor variants: original (fourfold), threefold, \twoFold{}, and threaded on our \textit{E.~coli} subsequences.}
    \label{fig:builderConstructors}
\end{figure}

\cref{fig:builderConstructors} shows the scalability of four variants of the dynamic index. 
We evaluate the original fourfold constructor as introduced in~\cite{osterkamp26cebwt}, our threefold variant, \twoFold{}, and the threaded variant, which also uses \twoFold{}. 
All tests use the same sequences of length 24{,}000 from before. The threaded constructor is shown with four worker threads as the threaded data point in this comparison; the effect of the number of workers is evaluated separately in \cref{section:genomedata}.

All four constructors exhibit near-linear growth in runtime. The threaded variant is the fastest throughout the experiment, reaching roughly 24{,}000--28{,}000 symbols per second after the first few sequences. \twoFold{} follows with roughly 13{,}000--17{,}000 symbols per second, while the threefold variant lies between \twoFold{} and the original fourfold construction. The fourfold variant is consistently the slowest, staying below about 9{,}000 symbols per second. For the first four sequences, the threaded variant shows a visible throughput increase, caused by the point at which enough independent partial-index construction work is available to compensate for the cost of merging.

For memory consumption, all variants need fewer bytes per symbol as more sequences are added.
The reduced and threaded variants use less memory than the fourfold construction for the first few sequence counts. As more sequences are added, the memory curves converge, and with 12 sequences all variants need almost the same amount of memory per symbol.

The runtime ordering is caused by the construction logic. The fourfold variant always processes almost twice the input \twoFold{}, 
directly increasing its cost. The advantage of the threaded variant over the \twoFold{} results from parallelizing the construction of partial indices. This parallelism does not remove the cost of merging the partial indices into the global dynamic index, and the thread-scaling experiment below shows that the ordered merge phase becomes the limiting factor once a small number of workers is available. The higher memory consumption for the first sequences is largely due to fixed overhead from the index and auxiliary structures, such as the conjugate array. As the input size increases, the observed decrease in bytes per symbol indicates that the additional memory overhead for construction does not grow proportionally, allowing the index to be built efficiently without significant extra memory from the index or auxiliary structures.

\begin{figure}[htbp]
    \centering
    \begin{subfigure}[b]{0.45\textwidth}
        \centering
        \includegraphics[width=\textwidth]{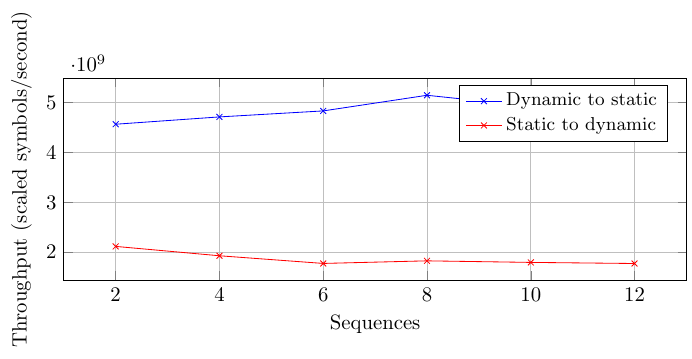}
        \caption{Throughput (symbols/second)}
        \label{fig:staticbuilderrun}
    \end{subfigure}
    \hfill
    \begin{subfigure}[b]{0.45\textwidth}
        \centering
        \includegraphics[width=\textwidth]{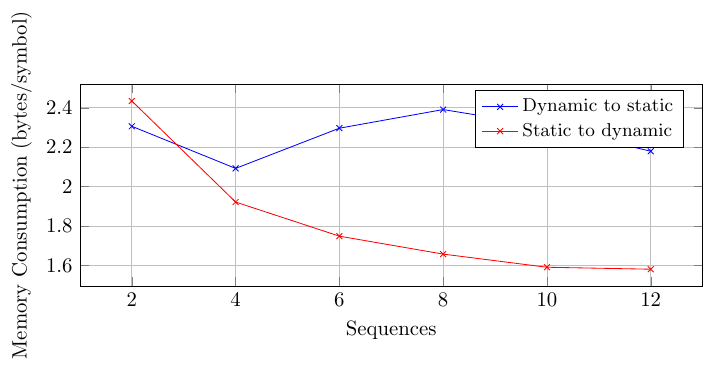}
        \caption{Memory consumption (bytes/symbol)}
        \label{fig:staticbuildermem}
    \end{subfigure}
    \caption{Comparison of throughput (left) and memory consumption per symbol (right) during dynamic-to-static and static-to-dynamic index transformation on our \textit{E.~coli} subsequences.}
    \label{fig:staticbuilder}
\end{figure}

\cref{fig:staticbuilder} shows throughput in symbols per second and peak memory consumption as bytes per symbol during conversion from the dynamic index to the static index and vice versa, using the same sequences as before.

Dynamic-to-static conversion achieves significantly higher throughput than static-to-dynamic conversion. Dynamic-to-static throughput grows slightly, exceeding 350{,}000 symbols per second for more sequences. Static-to-dynamic conversion remains stable at approximately 150{,}000 symbols per second.
For memory consumption, static-to-dynamic conversion steadily decreases from 2.4 to 1.6 bytes per symbol over 12 sequences. Dynamic-to-static conversion shows a slight increase between 4 and 8 sequences, peaking at 2.4 bytes per symbol, followed by a decline to 2.2 bytes per symbol. 

This behavior can be attributed to the nature of the transformation itself. The static index is derived directly from the existing dynamic index by traversing the $L$ and $F$ structures and converting integer values into binary unary representations across multiple bit vectors~\cite{osterkamp26cebwt}, resulting in $\mathcal{O}(n)$ time with minimal overhead. In contrast, dynamically constructing the wavelet tree one symbol at a time introduces significant runtime overhead compared to constructing all complete static bit vectors in a single pass. The irregularity between 4 and 8 sequences can be attributed to the inner loop iterating over alphabet values, whose distribution varies across sequences and can temporarily affect transformation cost. Static-to-dynamic conversion requires fewer auxiliary structures, as reconstructing the dynamic wavelet trees does not involve the same depth-dependent overhead and extra structures for temporarily storing values as constructing the bit vector collection. 

\begin{figure}[htbp]
    \centering
    \begin{subfigure}[b]{0.45\textwidth}
        \centering
        \includegraphics[width=\textwidth]{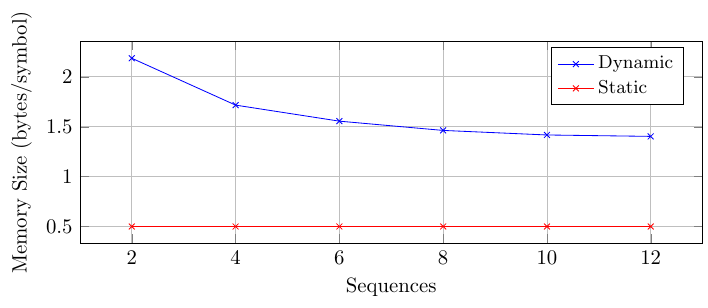}
        \caption{Main data structures}
        \label{fig:dataStructuremain}
    \end{subfigure}
    \hfill
    \begin{subfigure}[b]{0.45\textwidth}
        \centering
        \includegraphics[width=\textwidth]{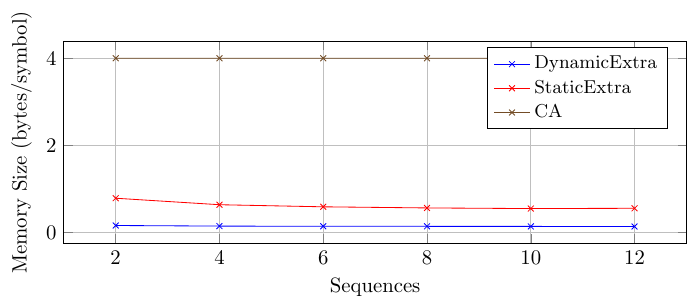}
        \caption{Additional data structures}
        \label{fig:dataStructureextra}
    \end{subfigure}
    \caption{Memory footprint of the main ($F$, $L$, $\mathrm{LCP}^{\infty}$; $B_L$, $B_F$) and additional ($e$; $\mathrm{LCP}^{\infty}$; $\mathrm{CA}$) data structures for the dynamic index, static index and conjugate array on our \textit{E.~coli} subsequences.}
    \label{fig:dataStructure}
\end{figure}

\cref{fig:dataStructure} shows the memory occupied by the index data structures themselves in bytes per symbol. We separate the main structures in \cref{fig:dataStructuremain} from the additional structures in \cref{fig:dataStructureextra}. For the static index, the main components include the bit vector collections $B_F$ and $B_L$. For the dynamic index, the $L$, $F$, and $\mathrm{LCP}^{\infty}$ wavelet trees constitute the main components. The additional structures consist of the bit vector $e$ for the dynamic index and the $\mathrm{LCP}^{\infty}$ wavelet tree of the dynamic index for the static index. The conjugate array is shown separately as it is shared between both index types.

The dynamic index requires more memory than the static index for its main structures. While the static index grows at a constant rate of 0.50 bytes per symbol, the dynamic index decreases from approximately 2.25 to less than 1.50 bytes per symbol before stabilizing.

For the additional structures, all three components exhibit an almost constant growth rate. The conjugate array grows at a constant rate of 4 bytes per symbol. The dynamic extra structures show a constant rate slightly over 0, while the static index extras decline slightly from 0.9 bytes per symbol to approximately 0.5 bytes per symbol at 12 sequences.

This behavior is a direct result of the underlying data structures. 
The higher memory usage of the dynamic index main structures reflects the overhead of dynamic wavelet trees compared to static bit vectors, as described in \cref{chapter:implementationdetails}. The dynamic index stores only the extra zero bit vector $e$, which is highly compact, whereas the static index retains the $\mathrm{LCP}^{\infty}$ wavelet tree of the dynamic index, which is considerably larger. The conjugate array stores full integer values in a plain vector, which was preferred over dynamic or static wavelet trees as they introduce a measurable time overhead during both construction and usage without a sufficient memory benefit.

\subsection{Internal Benchmarks on Genome Data} \label{section:genomedata}
\cref{tab:SortedBM} shows the throughput and memory consumption for three different sequence insertion orders based on sequence length: unordered, smallest first and largest first. The results are presented in tabular form to emphasize the key differences.

We observe that the smallest-first variant achieves the highest throughput with 28{,}199 symbols per second, followed by the unordered variant with 24{,}299 symbols per second. The largest-first variant has the lowest throughput with 23{,}360 symbols per second. In terms of memory consumption, the largest-first strategy requires the least memory per symbol with 34.32 bytes, followed by the unordered variant with 36.10 bytes and the smallest-first variant with 40.80 bytes per symbol.

The behavior can be explained by the interaction between sequence length, the incremental construction process of the dynamic index, and the threading model. When inserting sequences in increasing order of length, the construction begins with small, lightweight inputs. This allows workers to create multiple partial index structures that the merger can integrate efficiently. Because all intermediate structures must be stored temporarily, the resulting memory usage grows.

In contrast, inserting sequences in decreasing order of length keeps intermediate structures large during the early stages of construction. This reduces the number of concurrently maintained partial index structures and, therefore, reduces peak memory usage. However, when smaller sequences are inserted at later stages, they must be merged into already large index structures. This results in a higher number of costly comparison operations and more expensive updates, leading to increased runtime despite improved memory efficiency.

The unordered variant represents a balance between these two extremes. Since sequences of varying lengths are interleaved, neither large early-stage structures nor excessive late-stage extensions dominate the construction process. This results in a balanced overall runtime while maintaining moderate memory consumption compared to other strategies. 

This observation highlights an inherent trade-off in the design of the threaded construction. While the current approach allows multiple partial index structures to be constructed concurrently, thereby improving overall runtime through better utilization of available threads, it also increases peak memory consumption due to the temporary storage of these structures.
\begin{figure}[htbp]
  \centering
  \begin{subfigure}[b]{0.49\textwidth}
    \centering
    \begin{tabular}{|l|p{4cm}|}
        \hline
        \textbf{Ordering} & \textbf{Throughput (symbols/second)} \\
        \hline
        Unsorted & 24299 \\
Smallest First & 28199 \\
Biggest First & 23360 \\
       \end{tabular}\caption{Throughput}
    \label{tab:SortedRun}
  \end{subfigure}
  \hfill
  \begin{subfigure}[b]{0.49\textwidth}
    \centering
    \begin{tabular}{|l|p{4cm}|}
        \hline
        \textbf{Ordering} & \textbf{Memory consumption (bytes/symbol)} \\
        \hline
        Unsorted & 36.10 \\
Smallest First & 40.80 \\
Biggest First & 34.32 \\
       \end{tabular}\caption{Memory consumption}
    \label{tab:SortedMem}
  \end{subfigure}
  \caption{Comparison of throughput (left) and memory consumption per symbol (right) during dynamic index construction for different sequence insertion orders on 16 real genomes.}
  \label{tab:SortedBM}
\end{figure}

\begin{figure}[htbp]
    \centering
    \begin{subfigure}[b]{0.45\textwidth}
        \centering
        \includegraphics[width=\textwidth]{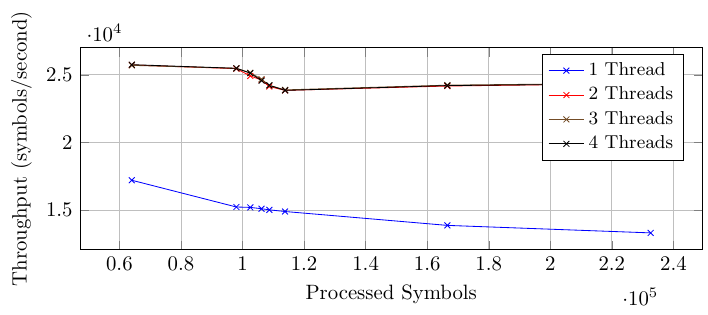}
        \caption{Runtime}
        \label{fig:ThreadedRum}
    \end{subfigure}
    \hfill
    \begin{subfigure}[b]{0.45\textwidth}
        \centering
        \includegraphics[width=\textwidth]{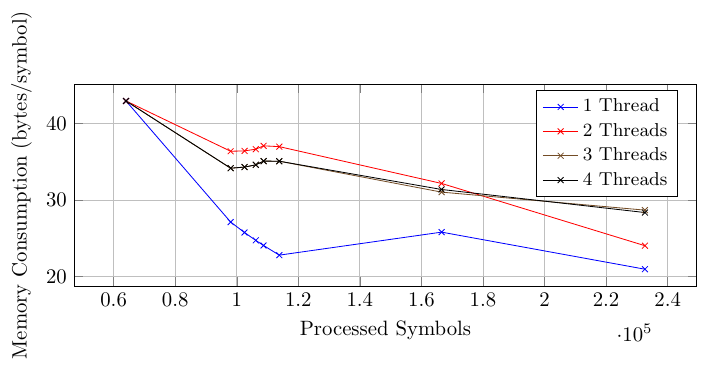}
        \caption{Memory consumption}
        \label{fig:ThreadedMem}
    \end{subfigure}
    \caption{Comparison of throughput (left) and memory consumption per symbol (right) during dynamic index construction with four different numbers of worker threads on 16 real genomes.}
    \label{fig:ThreadedBM}
\end{figure}

\cref{fig:ThreadedBM} shows the throughput and memory consumption for different numbers of worker threads. We evaluate configurations using one, two, three, and four workers.

The single-worker configuration is the slowest. Moving from one to two workers gives the clearest throughput improvement, whereas the three- and four-worker configurations do not provide a consistent additional gain. This indicates that the current threaded construction reaches its practical parallel limit after a small number of workers.

To distinguish lack of available parallel work from a scheduling artifact, we also profiled the threaded constructor by separating the time spent on partial-index construction, waiting for the next mergeable partial index, and merging into the global dynamic index. On the 16-genome benchmark with four workers, the summed partial-index construction time was about \(9.35\) seconds across worker threads, while the merger spent about \(2.15\) seconds waiting and \(5.63\) seconds merging. Since the partial time is summed over workers, the wall time is dominated by the ordered waiting and sequential merge phase. Approximately \(72\%\) of the merger-side time is spent in the actual merge and \(28\%\) in waiting for the next partial index in sequence order.

This behavior indicates that the scalability bottleneck is the ordered merge step rather than the construction of partial indices alone. Partial indices can be constructed concurrently, but they must be integrated into the global dynamic index in sequence order. Additional workers therefore increase the number of simultaneously materialized partial indices and the peak memory consumption, while the sequential merge phase remains largely unchanged. This behavior is consistent with Amdahl's law~\cite{amdahl_validity_1967}, but should be understood as a property of the present implementation rather than as a lower bound for all possible parallel ceBWT constructions.

The current implementation therefore favors deterministic ordered merging and reproducible benchmarking over aggressive out-of-order buffering. A more scalable design would have to reduce the cost of the merge itself or restructure the construction so that more of the merge work can be parallelized. Merely adding further worker threads is not sufficient for the present implementation. 

\begin{figure}[htbp]
    \centering
    \includegraphics[width=0.45\textwidth]{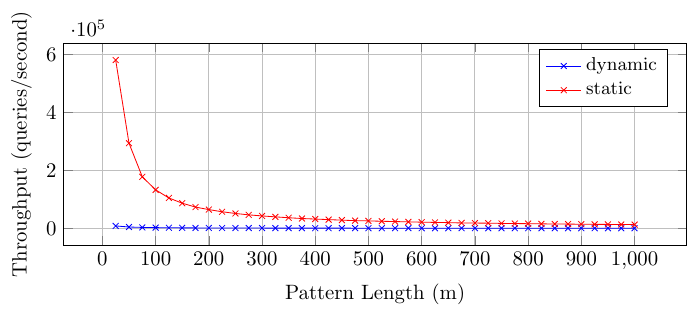}
    \caption{Scaling of the pattern search using the dynamic and the static index structure for different pattern lengths, measured on 50 patterns per length on 16 real genomes.}
    \label{fig:genomeeval}
\end{figure}

In \cref{fig:genomeeval} we show the scaling of pattern matching with growing pattern lengths.

Pattern search is faster with the static index structure. The dynamic index decreases from 8{,}000 to 238 queries per second, whereas the static index decreases from 500{,}000 to approximately 13{,}000 queries per second; both stabilize after a pattern length of 200.

The reason for the behavior lies in the underlying query structures used by each index, namely dynamic wavelet trees versus bit vectors, and the overall more compact search strategy of the static index, as explained in~\cite{osterkamp26cebwt}. 

\subsection{Internal Benchmarks on Repetitive Data} \label{section:internalrepet}
\begin{figure}[htbp]
    \centering
    \begin{subfigure}[b]{0.45\textwidth}
        \centering
        \includegraphics[width=\textwidth]{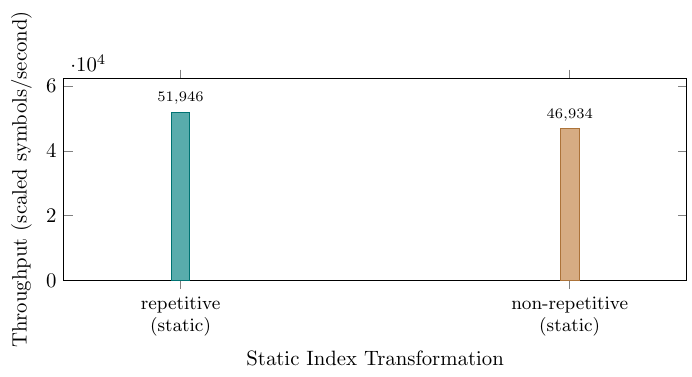}
        \caption{Throughput (symbols/second)}
        \label{fig:RepetitiveSBRuntime}
    \end{subfigure}
    \hfill
    \begin{subfigure}[b]{0.45\textwidth}
        \centering
        \includegraphics[width=\textwidth]{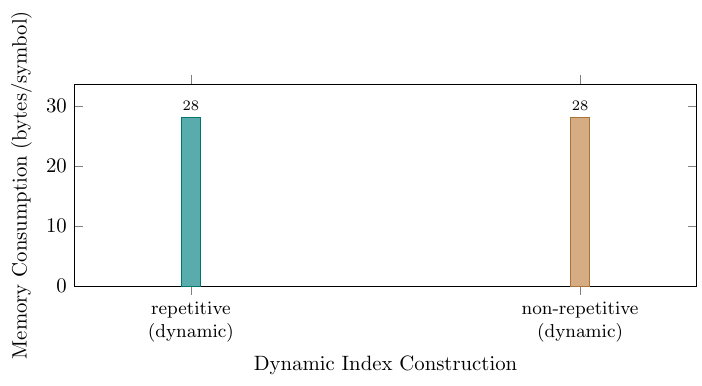}
        \caption{Memory consumption (bytes/symbol)}
        \label{fig:RepetitiveDBMemory}
    \end{subfigure}
    \caption{Comparison of throughput (left) during static index transformation and memory consumption per symbol (right) during dynamic index construction for 16 repetitive (Rice subsequences) versus non-repetitive genomes of identical length.}
    \label{fig:BuilderRepetitive}
\end{figure}

\cref{fig:BuilderRepetitive} presents the throughput of the static transformation and memory consumption during dynamic index construction for highly repetitive and non-repetitive sequences of identical length.

We focus on these specific plots since the memory consumption of the static transformation and throughput of the dynamic index construction were nearly identical across both datasets. Therefore, only the differing metrics are discussed in detail. Furthermore, the results are interpreted cautiously, as the experiments do not allow us to distinguish whether differences observed stem from structural properties of the sequences or from repetitiveness itself. Nevertheless, these results provide a first indication of how repetitiveness may influence index behavior in CTM and a more controlled investigation of this effect represents a promising direction for future work.

The repetitive dataset achieves a slightly higher throughput than the non-repetitive dataset, with a difference of 4{,}831 symbols per second. In contrast, memory consumption is the same at the displayed precision, with both datasets requiring 28 bytes per symbol.

Overall, the observed throughput difference is small, and the displayed memory consumption is identical. They are likely attributable to structural differences between the genomes rather than repetitiveness alone. In particular, since the peak heap usage reflects temporary allocations during construction, these variations are caused by differences in alphabet distributions and Cartesian tree signature value patterns, which influence intermediate computations without exhibiting a clear directional effect.

\begin{figure}[htbp]
    \centering
    \includegraphics[width=0.45\textwidth]{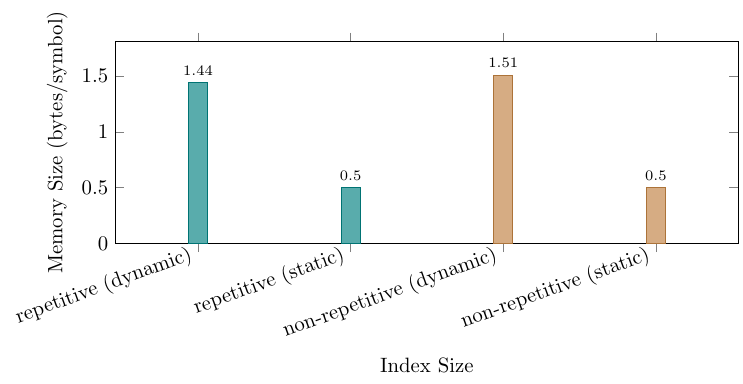}
    \caption{Memory footprint of the main static and dynamic index data structures for 16 repetitive versus non-repetitive genomes of identical length.}
    \label{fig:repetitive-data-structure}
\end{figure}

\cref{fig:repetitive-data-structure} shows the memory footprint of the index data structures for repetitive and non-repetitive datasets of identical length in bytes per symbol.

The repetitive dataset yields a slightly more compact dynamic index than the non-repetitive dataset, with a difference of 0.07 bytes per symbol. In contrast, the static index shows identical memory usage for both datasets.

The absence of any size difference for the static index results from the plain bit vector representation provided by SDSL, which does not benefit from repetitive structures under the memory bounds described in \cref{chapter:implementationdetails}. The behavior differs for the dynamic index. The observed difference is likely due to a more compact internal representation within the wavelet trees, as high repetitiveness leads to repeated value patterns that the underlying library can store more efficiently.

\begin{figure}[htbp]
    \centering
    \includegraphics[width=0.45\textwidth]{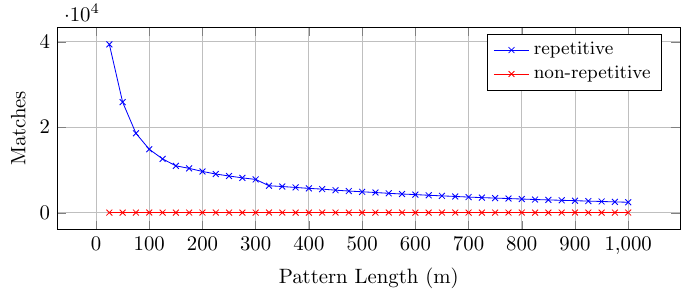}
    \caption{Number of matches found for repetitive versus non-repetitive genomes across varying pattern lengths.
    }
    \label{fig:repetitiveeval}
\end{figure}

A plot illustrating pattern matching runtime for both variants is available in the repository. However, because the two datasets use different pattern files and the runtime difference is negligible, we do not include it here. Instead, \cref{fig:repetitiveeval} shows how high repetitiveness influences the number of matches found.

Short patterns are found much more frequently in the repetitive dataset. However, longer patterns are also matched more often than in the non-repetitive dataset. Starting at nearly 40{,}000 matches for pattern length 25, the number of matches decreases sublinearly as the pattern length increases. For the non-repetitive dataset, it remains constant with each pattern occurring exactly once in the sequence, resulting in 50 matches.

This difference arises because repetitiveness significantly increases the probability that a randomly generated pattern occurs multiple times in the text. The fact that shorter patterns are found more frequently is a direct consequence of the structure of the input sequences. Although the patterns were generated randomly, the correlation between structural properties and pattern matching in CTM is clearly evident.

\subsection{Fitted Scaling Trends}  

The tables in this section compare the measured values with fitted asymptotic trend curves for construction time, index size, and pattern-search runtime. The fitted values are not independent theoretical predictions; they are least-squares fits~\cite[Section 15.4]{press_numerical_2007} of the asymptotic forms stated in the original ceBWT analysis~\cite{osterkamp26cebwt} and summarized in \cref{chapter:StateOfArt}. They are used only as a compact empirical check of whether the measured scaling behavior follows the expected asymptotic shape. For the first two experiments, we use the genome results from the threaded constructor from \cref{section:EColiBM}; for pattern search, we use the results from \cref{section:internalrepet}.

\cref{tab:TheoMeasBuilder} compares the fitted construction-time trend with the measured construction time of the dynamic builder as the sequence length increases. The asymptotic form used for the fit is $\mathcal{O}(n \frac{\lg\,\sigma\,\lg\,n}{\lg\,\lg\,n})$. 
The fitted curve stays close to the measurements over the tested range. This supports the conclusion that dynamic construction follows the expected scaling trend on these instances.

\begin{table}[t]
  \centering
  \begin{tabular}{|r|r|r|r|}
    \hline
    \textbf{Sequence Length (n)} & \textbf{Fit Dyn (s)} & \textbf{Dynamic (s)} & \textbf{\%} \\
    \hline
    48000 & 1.658 & 1.814 & 9\% \\
96000 & 3.451 & 3.468 & 1\% \\
144000 & 5.293 & 5.207 & -2\% \\
192000 & 7.168 & 7.112 & -1\% \\
240000 & 9.067 & 9.026 & -0\% \\
288000 & 10.985 & 11.069 & 1\% \\
   \end{tabular}\caption{Fitted asymptotic trend and measured construction time on the \textit{E.~coli} subsequences.}
  \label{tab:TheoMeasBuilder}
\end{table}

\cref{tab:TheoMeasData} compares the fitted memory trend with the measured space required to store the dynamic and static indexes. The asymptotic space bound for the dynamic data structure is $\mathcal{O}(n\,\lg\,\sigma)$ bits, and the static data structure uses $3n+o(n)$ bits.

The dynamic index is above the fitted trend for smaller instances and below it for the largest instances. This should not be interpreted as beating an asymptotic bound; rather, it reflects the hidden constants and lower-order effects captured by the fit over the tested range. The static index closely follows the fitted linear trend.

Since our dynamic wavelet trees scale with $S$ in $n(H_0 + 1)(1 + o(1)) + \mathcal{O}(|\Sigma| \log n)$ bits of space~\cite{prezza17dynamic} 
and since $H_0 \leq \lg\,\sigma$, the dynamic index is expected to satisfy the $\mathcal{O}(n\,\lg\,\sigma)$ bits of space
bound and may even perform better in practice when the empirical entropy is low. The static index uses bit vectors that occupy $n + o(n)$ bits of space~\cite{gog_sdsl_nodate}, which is consistent with the theoretical limit of $3n + o(n)$ bits in total. 
\begin{table}[htbp]
  \centering
  \resizebox{\textwidth}{!}{\begin{tabular}{|l|l|l|l|l|l|l|}
    \hline
    \textbf{Sequence Length (n)} & \textbf{Fit Dyn (KiB)} & \textbf{Dynamic (KiB)} & \textbf{\%} & \textbf{Fit Sta (KiB)} & \textbf{Static (KiB)} & \textbf{\%} \\
    \hline
    48000 & 68.217 & 102.500 & 50\% & 23.498 & 23.440 & -0\% \\
96000 & 136.433 & 160.960 & 18\% & 46.937 & 46.880 & -0\% \\
144000 & 204.650 & 218.820 & 7\% & 70.359 & 70.310 & -0\% \\
192000 & 272.866 & 274.340 & 1\% & 93.769 & 93.750 & -0\% \\
240000 & 341.083 & 332.230 & -3\% & 117.172 & 117.190 & 0\% \\
288000 & 409.300 & 394.720 & -4\% & 140.569 & 140.620 & 0\% \\
   \end{tabular}}
  \caption{Fitted asymptotic trend and measured memory size of the dynamic and static indexes on the \textit{E.~coli} subsequences.}
  \label{tab:TheoMeasData}
\end{table}

\cref{tab:TheoMeasEval} compares the fitted runtime trend with the measured runtime required to find patterns using the dynamic and static data structures. The asymptotic form for dynamic pattern search is $\mathcal{O}(m \frac{\lg\,\sigma\,\lg\,n}{\lg\,\lg\,n})$, and the asymptotic form for static pattern search is $\mathcal{O}(m)$.

The dynamic-index measurements stay close to the fitted trend. For the static index, the deviations vary between $7\%$ and $-8\%$ without a clear trend.

This behavior is directly related to the precomputation mentioned in~\cite[Theorem 3.16]{osterkamp26cebwt}. 
The key queries for backward search run in $\mathcal{O}(1)$ time, while the full computation takes $\mathcal{O}(m)$ time. The fluctuations with the static index are due to the overall fast times in the millisecond range.
\begin{table}[htbp]
  \centering
  \resizebox{\textwidth}{!}{\begin{tabular}{|l|l|l|l|l|l|l|}
    \hline
    \textbf{Pattern Length (m)} & \textbf{Fit Dyn (s)} & \textbf{Dynamic (s)} & \textbf{\%} & \textbf{Fit Sta (s)} & \textbf{Static (s)} & \textbf{\%} \\
    \hline
    100 & 0.0208 & 0.0219 & 5\% & 0.0004 & 0.0004 & 7\% \\
200 & 0.0416 & 0.0434 & 4\% & 0.0007 & 0.0007 & -6\% \\
300 & 0.0623 & 0.0640 & 3\% & 0.0011 & 0.0011 & -1\% \\
400 & 0.0831 & 0.0844 & 2\% & 0.0015 & 0.0015 & 1\% \\
500 & 0.1039 & 0.1049 & 1\% & 0.0019 & 0.0019 & 2\% \\
600 & 0.1247 & 0.1264 & 1\% & 0.0022 & 0.0023 & 3\% \\
700 & 0.1454 & 0.1456 & 0\% & 0.0026 & 0.0026 & -0\% \\
800 & 0.1662 & 0.1654 & -0\% & 0.0030 & 0.0030 & 1\% \\
900 & 0.1870 & 0.1858 & -1\% & 0.0033 & 0.0033 & -1\% \\
1000 & 0.2078 & 0.2063 & -1\% & 0.0037 & 0.0037 & -1\% \\
   \end{tabular}}
  \caption{Fitted asymptotic trend and measured runtime for dynamic and static pattern search on 16 real genomes.}
  \label{tab:TheoMeasEval}
\end{table}

In general, construction time, index size, and pattern-search runtime follow the fitted asymptotic trends on the evaluated instances.

\subsection{External Benchmarks}                   

\begin{figure}[htbp]
    \centering
    \begin{subfigure}[b]{0.48\textwidth}
        \centering
        \includegraphics[width=\textwidth]{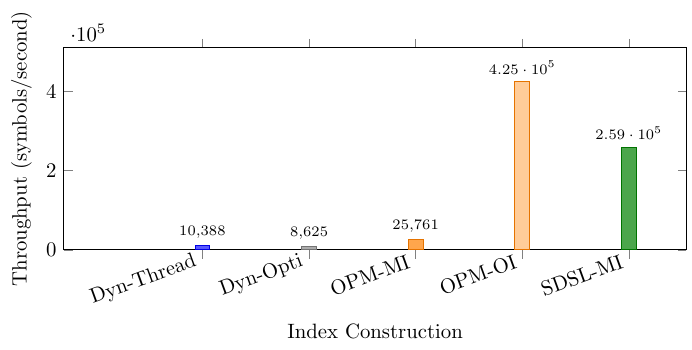}
        \caption{Throughput (symbols/second)}
        \label{fig:ExternBuildRun}
    \end{subfigure}
    \hfill
    \begin{subfigure}[b]{0.48\textwidth}
        \centering
        \includegraphics[width=\textwidth]{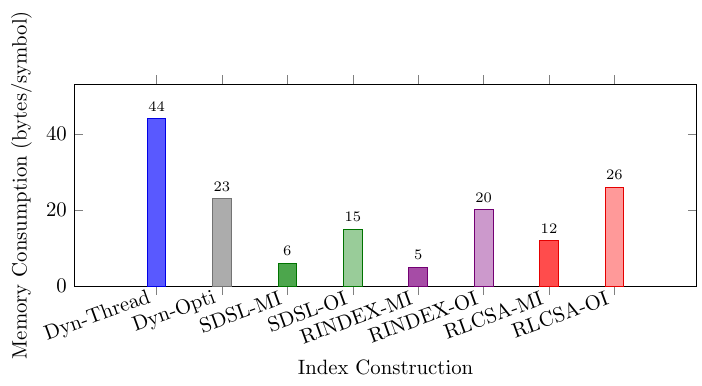}
        \caption{Memory consumption (bytes/symbol)}
        \label{fig:ExternBuildMem}
    \end{subfigure}
    \caption{Comparison of construction throughput (left) and memory consumption per symbol (right) on tokenized MIDI files. The throughput plot includes the exact string-matching baselines (SDSL, r-index, and RLCSA) and the OPM-FM index; the memory plot shows the implementations for which peak construction memory was measured.
        For visualization reasons, we omitted in the left figure those implementations for which construction throughput was orders of magnitude faster than our implementation.
    }
    \label{fig:externalbuilderbasic}
\end{figure}

\begin{figure}[htbp]
    \centering
    \begin{subfigure}[b]{0.48\textwidth}
        \centering
        \includegraphics[width=\textwidth]{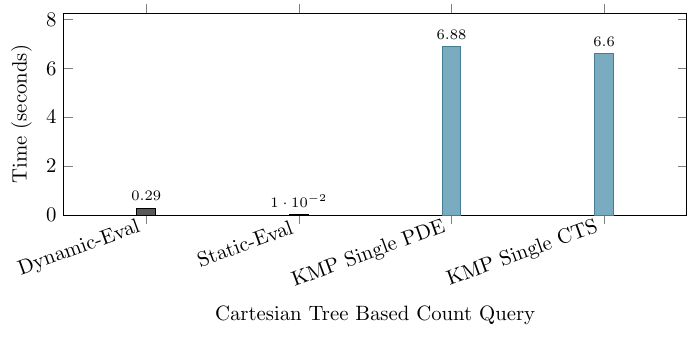}
        \caption{Runtime}
        \label{fig:kmprun}
    \end{subfigure}
    \hfill
    \begin{subfigure}[b]{0.48\textwidth}
        \centering
        \includegraphics[width=\textwidth]{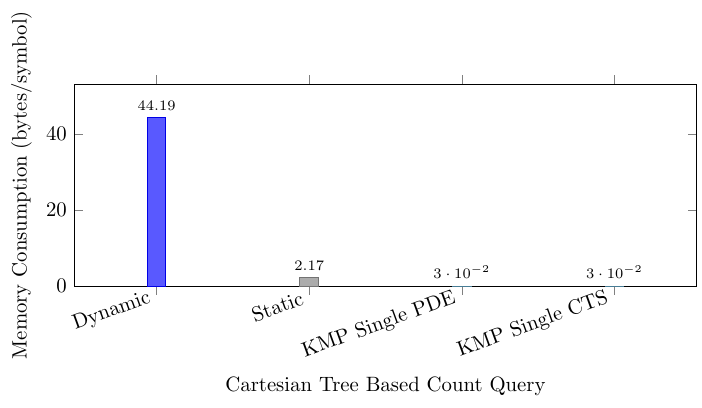}
        \caption{Memory consumption}
        \label{fig:kmpmem}
    \end{subfigure}
    \caption{Comparison of runtime (left) and memory consumption per symbol (right) between our ceBWT and KMP-based variants (PDE/CTS) on tokenized MIDI files for 2{,}000 patterns of different lengths. Runtime is measured for pattern search with an existing ceBWT index, whereas the ceBWT memory values report construction peak memory and the KMP memory values report ad hoc search memory.}
    \label{fig:kmp}
\end{figure}

\cref{fig:externalbuilderbasic} shows construction throughput (symbols/second) and peak memory consumption (bytes/symbol) during index construction. We compare our implementation with SDSL~\cite{gog14sdsl} using the FM-index with \texttt{csa\_wt}, the r-index~\cite{gagie18bwt,gagie20fully}, RLCSA~\cite{novak_rlcsa_2010}, and the compact OPM-FM index~\cite{decaroli19compact}. The exact string-matching baselines and the OPM baseline solve different matching problems and therefore serve as engineering reference points rather than semantically identical competitors. Comparisons with the KMP-based CTM baselines~\cite{park19cartesiantreematching} are provided in \cref{fig:kmp}. For our index construction, we show only the threaded variant and \twoFold{}, because the static index transformation introduces only negligible additional cost.
We write the suffix \texttt{-OI} in case that we build one index over the concatenation of the input texts,
and \texttt{-MI} in case that we build an index for each input text.
In the latter case, a count query answer is the sum of the count queries on each of the constructed indexes.

Our implementation requires significantly more time for index construction than the fastest baseline configurations. The threaded variant has a throughput of 10{,}342 symbols per second and \twoFold{} has 8{,}543 symbols per second. The OPM-FM baseline reaches 25{,}675 symbols per second in the multi-index configuration and 4.24\,$\cdot 10^5$ symbols per second in the one-index configuration, while the plotted SDSL multi-index configuration reaches 2.6\,$\cdot 10^5$ symbols per second.
In terms of peak memory consumption, \twoFold{} requires 23 bytes per symbol and our threaded variant 44 bytes per symbol. SDSL requires 6 bytes per symbol in the multi-index configuration and 15 bytes per symbol in the one-index configuration. The corresponding values are 5 and 20 bytes per symbol for the r-index, and 12 and 26 bytes per symbol for RLCSA. Construction-time memory for the OPM-FM index is not shown in this plot, because the current benchmark reports its construction throughput and final index size but not its peak construction memory.

Runtime differences are expected, as previous benchmarks have already indicated an increased construction cost for our implementation. \twoFold{}, the overhead of dynamic wavelet trees, and the additional complexity of storing CTM information all contribute to this increased runtime. The OPM-FM index also has to represent a structural matching relation rather than exact equality, but its construction is still substantially cheaper than dynamic ceBWT construction on this dataset. Memory consumption is driven by the same factors, with the threaded variant incurring additional overhead from concurrent index management when multiple index structures wait to be merged, as discussed in \cref{section:implementationdifferences}.

\begin{figure}[htbp]
    \centering
    \includegraphics[width=0.6\textwidth]{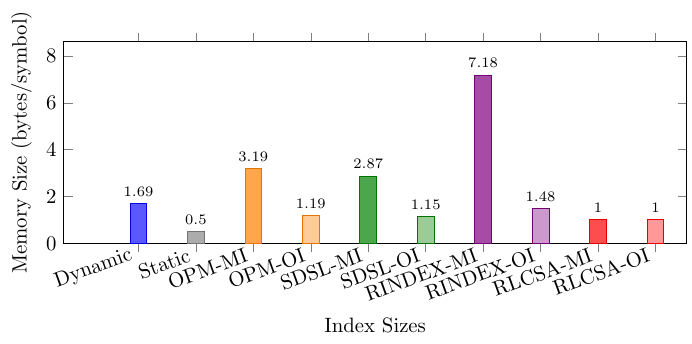}
    \caption{Memory-footprint comparison of the index structures among our ceBWT implementation, exact string-matching baselines (SDSL, r-index, and RLCSA), and the OPM-FM index on tokenized MIDI files.}
    \label{fig:ExternData}
\end{figure}

\cref{fig:ExternData} shows the memory footprint of the final index data structures for the same algorithms as before. For our implementation, we measure the main structures, since the additional components are not strictly required for pattern matching.
To determine the memory usage of our approach, we directly use the \texttt{bit\_size} methods provided by the respective libraries for wavelet trees and bit vectors. For SDSL and RLCSA, we also rely on the provided size methods. For the r-index, the occupied memory is determined using the \texttt{serialize} method and measuring the number of bytes written.

The static index achieves the smallest memory footprint with a total of 0.50 bytes per symbol, followed by RLCSA with 1.00 byte per symbol. The dynamic index uses 1.69 bytes per symbol, SDSL uses 2.87 bytes per symbol, and the OPM-FM index uses 3.19 bytes per symbol. The r-index is the largest in this experiment with 7.18 bytes per symbol.

The compact size of the static index can be attributed to the plain bit vector representation provided by SDSL, which stores the $B_F$ and $B_L$ structures with minimal overhead as described in \cref{section:implementationdifferences}. The dynamic index requires more space due to the overhead of dynamic wavelet trees, but still remains competitive with the exact and order-preserving indexes. The OPM-FM index is slightly larger than SDSL and our dynamic ceBWT index, but substantially smaller than the r-index on this multi-sequence MIDI dataset. The r-index is the largest due to the overhead of its run-length compressed representation and auxiliary structures~\cite{gagie18bwt,gagie20fully}, which are required for every one of the 203 sequences. RLCSA achieves a compact representation through compressed suffix arrays with less overhead information, but at the cost of slower query performance~\cite{novak_rlcsa_2010,siren09rlcsa}. The higher memory usage of the SDSL implementation can be attributed to the use of the \texttt{csa\_wt} structure, which is based on wavelet trees, as described in~\cite{gog14sdsl,makinen05rle}. This representation introduces additional space overhead to support efficient rank and select operations.

\begin{figure}[htbp]
    \centering
    \includegraphics[width=0.6\textwidth]{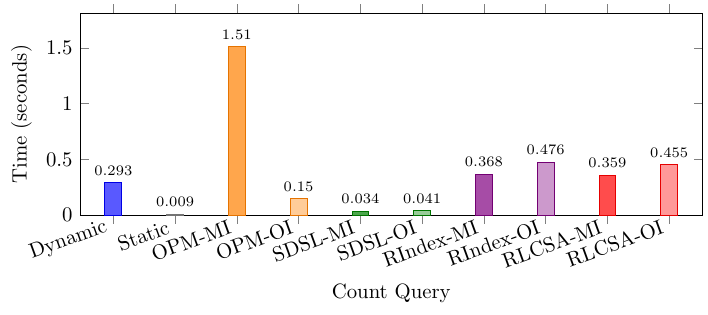}
    \caption{Pattern-search runtime comparison among our ceBWT implementation, exact string-matching baselines (SDSL, r-index, and RLCSA), and the OPM-FM index on tokenized MIDI files for 2{,}000 patterns of different lengths.}
    \label{fig:ExternEval}
\end{figure}

\cref{fig:ExternEval} compares the pattern matching performance of our implementation with the exact string-matching baselines and the OPM-FM index.
We search for a collection of 2{,}000 patterns of varying lengths in all sequences. 

The exact SDSL baselines achieve the fastest pattern matching, taking 0.034 seconds in the multi-index configuration and 0.041 seconds in the one-index configuration. Our static index follows with 0.084 seconds and is faster than both OPM-FM configurations, which take 1.512 seconds and 0.150 seconds, respectively. The r-index and RLCSA take between 0.358 and 0.470 seconds depending on the configuration, while the dynamic ceBWT index is the slowest in this experiment with 3.239 seconds.

Pattern-matching performance is partially correlated with the memory footprint of each index. The static index benefits from $\mathcal{O}(1)$ rank and select queries on plain SDSL bit vectors, which enable fast backward search with minimal overhead, as described in \cref{section:implementationdifferences}. SDSL achieves the lowest query time through its wavelet tree-based CSA representation, which also supports efficient rank and select operations~\cite{gog14sdsl}. RLCSA and the r-index are slower despite their compressed representations, because their query operations are more complex and incur additional overhead on this multi-sequence workload~\cite{novak_rlcsa_2010,siren09rlcsa,gagie18bwt,gagie20fully}. The dynamic ceBWT index is substantially slower than the static ceBWT index due to the slower rank and select queries of the dynamic wavelet trees and the more complex backward-search operations, as already observed in the internal benchmarks in \cref{section:internalrepet}. The OPM-FM index solves a structural matching problem different from exact matching and, in this benchmark, its query time is higher than the static ceBWT but lower than the dynamic ceBWT.

\cref{fig:kmp} compares our ceBWT indexes with two KMP-based CTM baselines following Park et al.~\cite{park19cartesiantreematching}. The first baseline uses PDE, whereas the second uses CTS together with the auxiliary array $D$ described in \cref{section:CTS}. Both baselines perform an online scan and do not construct a persistent index. We therefore use a different evaluation setup: we focus on the pattern-matching phase using already constructed ceBWT index structures while also reporting peak memory consumption for fairness.

For both patterns and sequences, we use a preloaded \texttt{std::vector<int>} representation, because this is more appropriate here with more than 2{,}000 pattern files and 203 sequence files. In such scenarios, file I/O overhead dominates the actual computation time.

With an existing ceBWT index, both index variants answer the full collection of pattern queries faster than the KMP-based baselines: the dynamic and static ceBWT variants take 3.24 and 0.08 seconds, respectively, whereas KMP takes 7.01 seconds with PDE and 6.82 seconds with CTS. The memory numbers in \cref{fig:kmp} measure different phases and should therefore be interpreted separately. The KMP variants allocate 34.16 bytes per symbol during ad hoc search, while the ceBWT numbers report construction peak memory, with 44.02 bytes per symbol for the dynamic builder and 1.82 bytes per symbol for the static builder.

The KMP-based approach avoids constructing a persistent index, but it still allocates pattern- and text-dependent working data during ad hoc search. In this experiment, this working memory is lower than the peak memory of the dynamic ceBWT construction but much higher than the peak memory reported for the static ceBWT construction. When using an existing ceBWT index with multiple sequences stored, pattern searches are faster because they operate directly on the prebuilt index, whereas the KMP-based approach must compare the pattern against sequences for every search. Therefore, the ceBWT is well-suited for long-term storage and repeated pattern searches, while the KMP approach is more appropriate for index-free, ad hoc searches on new datasets.

\begin{figure}[htbp]
    \centering
    \includegraphics[width=0.45\textwidth]{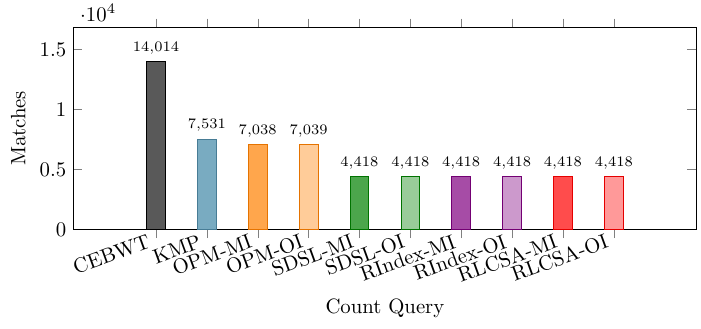}
    \caption{Number of matches found for each algorithm on tokenized MIDI files for 2{,}000 patterns of varying lengths.}
    \label{fig:ExternMatches}
\end{figure}

\cref{fig:ExternMatches} shows the total number of matches found in all tested implementations. 
We collate all variants of our ceBWT implementation and of the KMP-based variants. 
Since both solve count queries for circular CTM, we observe the same number of matches.
The difference between the exact string-matching baselines, OPM, and CTM arises from the generalized matching definitions itself. 
Exact matching requires equality of symbols, OPM abstracts from absolute values and preserves only their relative order, and CTM further abstracts to the Cartesian tree structure as described in \cref{section:CTM}. 
This explains why OPM finds more matches than exact string matching, while the CTM baselines find slightly more matches than OPM on this dataset. 
Further, non-circular matching has at most as many matches as circular matching,
which searches not only the original sequences but also all their rotations, as described in \cref{section:CEBWT}.

\section{Conclusion}\label{sec:conclusion-outlook}
This article evaluates the Cartesian Extended Burrows--Wheeler Transform (ceBWT) using a custom implementation on genomic and MIDI datasets. The objective was to analyze runtime, memory consumption, and scalability of a BWT-based CTM algorithm and to compare it with classical exact index structures, an order-preserving FM-index, and an alternative CTM approach. The research questions posed in \cref{chapter:introduction} can be answered as follows.

Index construction follows the expected fitted scaling trend on the evaluated instances,
with throughput decreasing with increasing sequence length due to the overhead of dynamic extension and the underlying dynamic wavelet trees. Computing and inserting the values required for dynamic extension constitute a major bottleneck, limiting practical scalability to smaller sequences.
Memory consumption follows a similar pattern, with the dynamic index requiring more space than the static index due to the overhead of dynamic wavelet trees.

The static index outperforms the dynamic index in both pattern search speed and storage efficiency and achieves index sizes competitive with the exact and order-preserving baselines. On the MIDI dataset, it is also faster at query time than the compact OPM-FM index~\cite{decaroli19compact}, although OPM-FM is cheaper to build and solves a different matching problem. Compared with the two KMP-based CTM variants~\cite{park19cartesiantreematching}, the results show that a maintained index structure can substantially accelerate repeated pattern searches, at the cost of index construction and stored index space.

Repetitiveness has a measurable but modest effect on memory consumption and construction speed, although these differences are difficult to attribute to repetitiveness alone. However, it has a significant impact on the number of matches found. Overall, OPM and CTM both produce more matches than exact string-matching baselines, as they consider structural equivalences rather than exact values. CTM produces slightly more matches than OPM on the MIDI benchmark. The ceBWT and the KMP-based CTM baselines report the same circular CTM match counts in our setup.

Implementation challenges arose primarily from the dynamic extension mechanism, the conjugate array ordering, and the interaction between parallel construction and measurement instrumentation. The threaded constructor requires deterministic sequencing of partial indices and a single-threaded fallback for very small inputs. Profiling shows that its scalability is limited primarily by ordered merging into the global dynamic index: additional workers can construct partial indices concurrently, but the resulting partial indices still have to be merged sequentially. Moreover, heap-memory measurements for threaded construction require thread-safe allocator instrumentation, since allocator hooks are active while worker threads are created. 

In conclusion, this work presents a first systematic empirical analysis of a ceBWT-based CTM implementation.
It identifies advantages such as efficient storage and high match rates, as well as limitations, particularly in dynamic construction. Although the current implementation is limited to smaller sequences, the experiments demonstrate the practical applicability of the approach to real-world data and provide a solid foundation for future optimizations and research.

\section{Outlook}
Based on the results, several promising directions for future research can be identified:
\begin{itemize}
    \item Optimizing dynamic structures: The overhead of dynamic extension can be reduced through more efficient rank and select queries. Additionally, improved insert and remove operations could significantly enhance runtime performance, particularly for larger sequences.
    \item Exploring new data domains: Applying the approach to sequences with domain-specific alphabets, such as financial or sensor data, could extend its applicability in pattern recognition.
    \item Designing hybrid index systems: Combining static and dynamic index structures may enable flexible usage in both analytical and real-time scenarios.
    \item Studying scalability and repetitiveness: Further investigation into the effect of repetitive structures on memory consumption and match frequency could lead to targeted optimizations for highly repetitive datasets.
\end{itemize}

\bibliographystyle{plain}

\clearpage

\appendix
\section{Additional Figures and Tables}

\begin{longtable}{|p{0.18\textwidth}|p{0.47\textwidth}|p{0.26\textwidth}|}
\caption{Notation quick reference. The last column indicates where the symbol is first introduced or first used with this meaning.}
\label{tab:notation-quick-reference}\\
\hline
\textbf{Notation} & \textbf{Meaning} & \textbf{First appearance} \\
\hline
\endfirsthead
\hline
\textbf{Notation} & \textbf{Meaning} & \textbf{First appearance} \\
\hline
\endhead
\hline
\endfoot
$i,j,k,q$ & Integer indices; $q$ denotes a global text position in the concatenated indexed collection. & opening paragraph; \cref{sec:ceBWT-query} \\
\hline
$[i..j]$ & Integer interval $\{i,i+1,\ldots,j\}$, empty if $j<i$. & opening paragraph \\
\hline
$X,Y,Z,U,W,V$ & Generic strings or comparison strings. & opening paragraph; \cref{sec:omega-order} \\
\hline
$S$ & Sequence used in examples or as a single input string. & \cref{sec:ctm-definitions} \\
\hline
$T,T_j$ & Indexed text; $T_j$ is the $j$-th text in a collection. & \cref{sec:ceBWT-query} \\
\hline
$P$ & Query pattern. & \cref{sec:ctm-definitions}; \cref{sec:ceBWT-query} \\
\hline
$m$ & Pattern length in complexity statements; also used as a local string length in construction examples. & \cref{chapter:StateOfArt}; \cref{sec:construction-model} \\
\hline
$|X|$, $X[i]$, $X[i..j]$, $X^\omega$ & Length, symbol access, substring, and infinite periodic repetition of $X$. & opening paragraph \\
\hline
$\Sigma$, $\sigma$ & Finite totally ordered alphabet and its size; when convenient, $\Sigma$ is identified with $[1..\sigma]$. & opening paragraph \\
\hline
$\$$ & Separator symbol not in $\Sigma$, smaller than all alphabet symbols. & opening paragraph \\
\hline
$\infty$ & Symbol larger than every integer; used by PDEs. & \cref{section:CTS} \\
\hline
$\ctree{X}$ & Cartesian tree of $X$. & \cref{sec:ctm-definitions} \\
\hline
$X\ctmatch Y$ & Cartesian-tree matching relation; $X$ and $Y$ have identical Cartesian trees. & \cref{sec:ctm-definitions} \\
\hline
$\pde{X}$ & Parent-distance encoding of $X$. & \cref{section:CTS} \\
\hline
$\lambda_S$ & Cartesian tree signature of a single sequence or current KMP window $S$. & \cref{section:CTS} \\
\hline
$D$ & Auxiliary array used with $\lambda_S$ for CTS-based KMP window updates. & \cref{section:CTS} \\
\hline
$\rot{X}{k}$ & $k$-th cyclic rotation of $X$; also called a conjugate. & \cref{sec:ceBWT-query} \\
\hline
$\mathcal T=(T_1,\ldots,T_d)$ & Indexed collection of $d$ texts. & \cref{sec:ceBWT-query} \\
\hline
$n$, $n_k$, $d$ & Total number of indexed symbols, length of $T_k$, and number of indexed texts. & \cref{sec:ceBWT-query} \\
\hline
$\conj{\mathcal T}{q}$ & Conjugate starting at global position $q$ in the indexed collection. & \cref{sec:ceBWT-query} \\
\hline
$\operatorname{count}(P)$ & Number of circular CTM occurrences of $P$ in the indexed collection. & \cref{sec:ceBWT-query} \\
\hline
$\varepsilon$ & Empty pattern; its conjugate range is $[1..n]$ when allowed. & \cref{sec:omega-order} \\
\hline
$\rpde{X}$ & Rotational PDE of $X$, defined as $\pde{X^2}[|X|+1..2|X|]$. & \cref{sec:omega-order} \\
\hline
$\primitive{Z}$ & Primitive root of a nonempty string $Z$. & \cref{sec:omega-order} \\
\hline
$\ctpeq$, $\cteq$, $\ctprec$ & $\omega$-preorder, equivalence, and strict order of conjugates. & \cref{sec:omega-order} \\
\hline
$\CA{\mathcal T}$, $\ICA{\mathcal T}$ & Conjugate array and its inverse. & \cref{sec:omega-order}; \cref{sec:lf-definition} \\
\hline
$\crange{\mathcal T}{P}=[\ell..r]$ & Conjugate-array interval containing the circular CTM matches of $P$. & \cref{sec:omega-order} \\
\hline
$\operatorname{prev}_{\mathcal T}(q)$ & Previous text position in the same text cycle, with periodic-equivalence handling. & \cref{sec:lf-definition} \\
\hline
$\LF{\mathcal T}$, $\FL{\mathcal T}$ & LF mapping and its inverse over the ordered conjugates. & \cref{sec:lf-definition} \\
\hline
$\rts{X}$, $\pi(X)$ & Rotational Cartesian tree signature and its first value $\pi(X)=\rts{X}[1]$. & \cref{sec:lf-definition} \\
\hline
$\Farr{\mathcal T}$, $\Larr{\mathcal T}$ & Stored ceBWT arrays that replace explicit $\LF{\mathcal T}$/$\FL{\mathcal T}$ storage. & \cref{sec:lf-definition} \\
\hline
$\rankop_c(A,i)$, $\selectop_c(A,t)$ & Rank and select operations for value $c$ in array or bit-vector representation $A$. & \cref{sec:lf-definition} \\
\hline
$\LCPinf{\mathcal T}$, $\lcpct{X}{Y}$ & LCP array for adjacent conjugates and the corresponding CTM-specific LCP measure. & \cref{sec:index-variants} \\
\hline
$e$, $h$ & Pattern-side backward-search values, with $h=\pi(P[i..|P|]\$)$ and $e$ counting relevant $\infty$ values in the PDE suffix. & \cref{sec:index-variants} \\
\hline
$E$ & Auxiliary dynamic bit vector used during incremental construction. & \cref{sec:index-variants} \\
\hline
$B_{\Farr{\mathcal T}}^c$, $B_{\Larr{\mathcal T}}^c$; $B_F$, $B_L$ & Static bit-vector encodings of the $F$ and $L$ arrays; $B_F$ and $B_L$ denote the implemented collections. & \cref{sec:index-variants} \\
\hline
$R=T^4\$$ & Fourfold replicated construction text with a separator. & \cref{sec:construction-model} \\
\hline
$\rta{X}{i}$ & $i$-th left rotation of $X$ in construction examples; same conjugate as $\rot{X}{i}$. & \cref{sec:construction-model} \\
\hline
$\mathcal R$, $\rho$ & Already indexed collection during insertion and its number of conjugates. & \cref{sec:construction-model} \\
\hline
$\cnt{\mathcal R}{V}$ & Insertion rank of comparison string $V$ among the old conjugates of $\mathcal R$. & \cref{sec:construction-model} \\
\hline
$\plcp{\mathcal R}{V}$, $\slcp{\mathcal R}{V}$ & Predecessor and successor LCP helper values used during merging. & \cref{sec:construction-model} \\
\hline
$\cntarr{i}$, $\plcparr{i}$, $\slcparr{i}$ & Table abbreviations for the count, predecessor LCP, and successor LCP values of rotation $\rta{S}{i}$. & \cref{sec:construction-model} \\
\hline
$H_0$ & Zero-order entropy of a sequence, used when quoting dynamic wavelet-tree space bounds. & \cref{chapter:implementationdetails} \\
\hline
\end{longtable}

\end{document}